\documentclass[12pt,letterpaper]{JHEP3}
\usepackage{latexsym,amsmath,amsfonts,amssymb}
\usepackage{graphicx}
\usepackage{bbm}
\usepackage{multirow}

\usepackage{color}

\newcommand\trans{{\ensuremath{\text{\sf T}}}}

\newcommand{\eg}{\textit{e.g.}}

\newcommand{\ie}{\textit{i.e.}}

\numberwithin{equation}{section}

\newcommand{\be}{\begin{equation}} \newcommand{\ee}{\end{equation}}
\newcommand{\bea}{\begin{equation} \begin{aligned}} \newcommand{\eea}{\end{aligned} \end{equation}}
\newcommand{\tabs}{\rule[-1ex]{0pt}{4ex}}

\newcommand{\cA}{\mathcal{A}}
\newcommand{\cB}{\mathcal{B}}
\newcommand{\cC}{\mathcal{C}}

\newcommand{\cM}{\mathcal{M}}
\newcommand{\cN}{\mathcal{N}}

\newcommand{\cS}{\mathcal{S}}

\newcommand{\bC}{\mathbb{C}}

\newcommand{\bR}{\mathbb{R}}
\newcommand{\bZ}{\mathbb{Z}}
\newcommand{\fg}{\mathfrak{g}}
\newcommand{\fh}{\mathfrak{h}}

\def\SO{\mathrm{SO}}
\def\O{\mathrm{O}}
\def\SU{\mathrm{SU}}
\def\U{\mathrm{U}}

\def\USp{\mathrm{USp}}

\def\su{\mathfrak{su}}
\def\u{\mathfrak{u}}
\def\usp{\mathfrak{usp}}

\def\so{\mathfrak{so}}

\DeclareMathOperator{\tr}{tr}
\DeclareMathOperator{\rk}{rk}

\def\hkq{\big/\!\!\big/\!\!\big/}
\def\Otm#1{\raisebox{0pt}[0pt][0pt]{$\widetilde{\text{O#1}}\phantom{I\!\!\!}^-$}}
\def\Otp#1{\raisebox{0pt}[0pt][0pt]{$\widetilde{\text{O#1}}\phantom{I\!\!\!}^+$}}

\title{Mirrors of 3d Sicilian theories}

\let\AA\dagger
\let\BB\ddagger
\let\CC\S
\author{Francesco Benini$^\AA$, Yuji Tachikawa$^\BB$, and Dan Xie$^\CC$\\

{\tt fbenini@princeton.edu, yujitach@ias.edu, fogman@tamu.edu}\\
$^\AA$ Department of Physics, Princeton University, \\
Princeton, NJ 08544, USA \\
$^\BB$ School of Natural Sciences, Institute for Advanced Study, \\
Princeton, NJ 08540, USA \\
$^\CC$ George P.~and Cynthia W.~Mitchell Institute for Fundamental Physics, \\
Texas A\&M University, College Station, TX 77843, USA.
}

\preprint{MIFPA-10-27\\
PUTP-2344}

\abstract{%
We consider the compactification of the 6d $\cN=(2,0)$ theories,
or equivalently of M-theory 5-branes, on a punctured Riemann surface times a circle.
This gives rise to what we call 3d $\cN=4$ Sicilian theories,
and we find that their mirror theories are star-shaped quiver gauge theories.
We also discuss an alternative construction of these 3d theories through 4d $\cN=4$ SYM on a graph, which allows us to obtain the 3d mirror via 4d S-duality.
}

\keywords{3d mirror symmetry, Sicilian theory}

\begin{document}

\section{Introduction}

Last year Gaiotto showed in \cite{Gaiotto:2009we} that the class of 4d $\cN=2$ theories which naturally arise is far larger than was thought before, by extending an observation by Argyres and Seiberg \cite{Argyres:2007cn,Argyres:2007tq}.
Namely, it was argued that the strong-coupling limit of $\cN=2$ superconformal gauge theories coupled to hypermultiplets almost always involves a plethora of newly-discovered non-Lagrangian theories.%
\footnote{In this paper, a theory is called Lagrangian if there is a UV Lagrangian consisting of vector, spinor and scalar fields whose IR limit equals that theory. We call a theory non-Lagrangian when such a description is not known. We find this definition useful, although it is admittedly time-dependent.  With this definition, the $E_6$ theory of Minahan and Nemeschansky is non-Lagrangian: 4d $\SU(3)$ theory with six flavors in the strong coupling limit consists of the $E_6$ theory with $\SU(2)$ gauge field and hypermultiplets \cite{Argyres:2007cn}, but we do not yet have a Lagrangian which realizes only the $E_6$ theory.  }
A most prominent of them is the so-called $T_N$ theory, which has $\SU(N)^3$ flavor symmetry.

These  theories describe the low-energy dynamics of $N$ M5-branes wrapped on a Riemann surface with punctures. We call this class the \emph{Sicilian theories} of type $A_{N-1}$.
Each puncture is associated with a Young diagram specifying the behavior of the worldvolume fields at the puncture.
Any Riemann surface with punctures can be constructed by taking a number of spheres with three punctures, and connecting pairs of punctures by tubes.
Correspondingly, a general Sicilian theory can be constructed by taking a number of \emph{triskelions},
which are the low-energy limit of $N$ M5-branes on a sphere with three punctures,
and gauging together their flavor symmetries by gauge multiplets.
The most important among the triskelions is the $T_N$ theory, from which all other triskelions can be generated by moving along the Higgs branch.

Further generalizations have been pursued in many directions, \eg{} their gravity dual \cite{Gaiotto:2009gz} and a IIB brane realization \cite{Benini:2009gi} have been found, their superconformal indices have been studied \cite{Gadde:2009kb,Gadde:2010te}, the theories have been extended to the case with orientifolds \cite{Tachikawa:2009rb, Nanopoulos:2009xe}, or lower $\cN=1$ supersymmetry \cite{Maruyoshi:2009uk,Benini:2009mz}, etc.
In this paper we study the properties of these theories by compactifying them on $S^1$, which leads to 3d $\cN=4$ superconformal theories when the radius is made small.

The vacuum moduli space of a 3d $\cN=4$ theory has a Coulomb branch and a Higgs branch. Fluctuations on the first are described by massless vector multiplets and fluctuations on the latter by massless hypermultiplets.
However there is no fundamental distinction between the two types of multiplets.
Namely, they only differ by the assignment of the representation of the R-symmetry $\SO(4)_R \simeq \SO(3)_X \times \SO(3)_Y$,
and the exchange of $\SO(3)_X$ and $\SO(3)_Y$ maps massless vector multiplets into massless hypermultiplets and vice versa.
Therefore, for a theory $\cA$ there is another theory $\cB$ such that the Coulomb branch of $\cA$ is the Higgs branch of $\cB$ and vice versa.  $\cB$ is called the \textit{mirror} of $\cA$, and this operation is called 3d \textit{mirror symmetry} \cite{Intriligator:1996ex}.

\begin{figure}
\[
\vcenter{\hbox{\includegraphics[height=5cm]{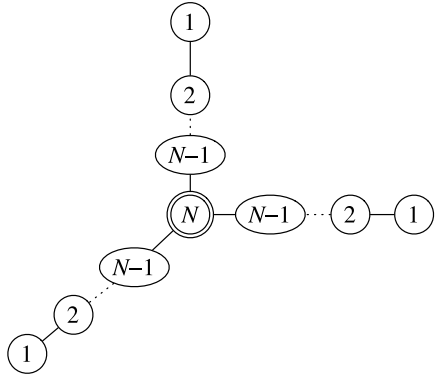}}}
\]
\caption{The mirror of the 3d $T_N$ theory. A circle with $n$ inside stands for a $\U(n)$ gauge group, a doubled circle an $\SU(n)$, and a line between two circles signifies a bifundamental hypermultiplet transforming under the two gauge groups connected by the line. \label{TN}}
\end{figure}

It is not guaranteed that the mirror of a Lagrangian theory is given by a Lagrangian theory. Indeed in \cite{Intriligator:1996ex} it was found that the mirrors of quiver theories based on the extended Dynkin diagrams of $E_{6,7,8}$ are non-Lagrangian theories with $E_{6,7,8}$ flavor symmetry.
These 3d  theories arise from the $S^1$ compactification of similar 4d non-Lagrangian theories found by Minahan and Nemeschansky \cite{Minahan:1996fg,Minahan:1996cj},
which are, in turn, prototypical examples of Sicilian theories.
We will see that the mirror of a Sicilian theory always has a Lagrangian description: it is a quiver gauge theory.
For example, the mirror of the 3d $T_N$ theory is given by a quiver gauge theory of the form shown in figure~\ref{TN}.
There, a circle with a number $n$ inside stands for a $\U(n)$  gauge group, a double circle for an $\SU(n)$ gauge group, and a line between two circles corresponds to a bifundamental hypermultiplet for the two gauge groups connected.
When $N=3$, the $\SU(3)^3$ symmetry of $T_3$ is known to enhance to $E_6$, and the quiver shown in figure~\ref{TN} indeed is the extended Dynkin diagram of $E_6$, reproducing the classic example in \cite{Intriligator:1996ex}.
In general, the mirror of the Sicilian theory for the sphere with $k$ punctures is a star-shaped quiver with $k$ arms coupled to the central node $\SU(N)$.\footnote{General star-shaped quivers have been studied mathematically \eg~by Crawley-Boevey \cite{CB}, and their relation to the moduli space of the Hitchin system which underlies the Coulomb branch of Sicilian theories were known to mathematicians, see \eg~\cite{Boalch}.}

\begin{figure}
\[
\includegraphics[width=.88\textwidth]{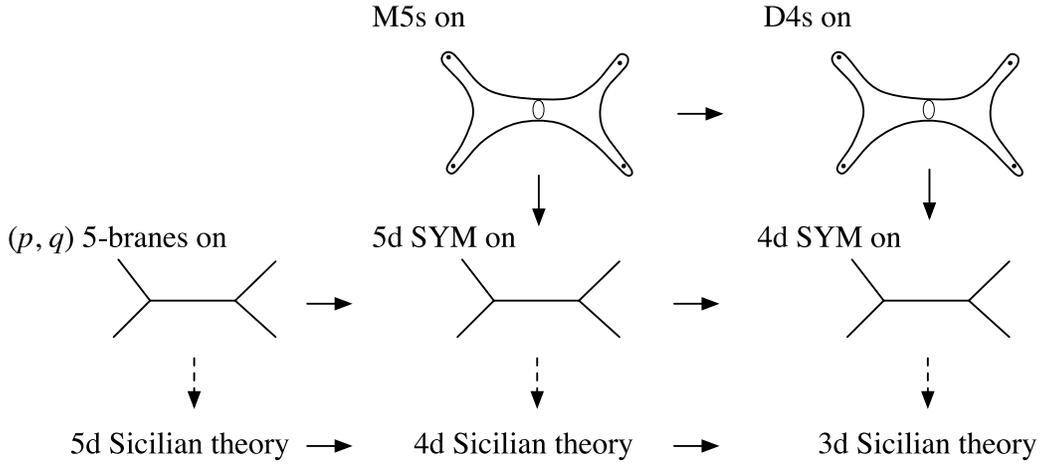}
\]
\vspace{-.7cm}
\caption{Relations of various constructions we will use in the paper. The horizontal arrow designates Kaluza-Klein reduction on $S^1$. The solid vertical arrow is a reduction on an $S^1$ inside the Riemann surface. The dotted vertical arrow signifies taking the long-distance limit compared to the size of the graph. \label{construction}}
\end{figure}

We will arrive at this observation via a brane construction, which is summarized in figure \ref{construction} and will be explained in the paper.  The mirrors of standard gauge theories were derived using D3-branes suspended between 5-branes in \cite{Hanany:1996ie}. This was then phrased in terms of a web of 5-branes compactified on $T^2$ in \cite{Aharony:1997ju}.
This latter method is readily applicable to the construction of Sicilian theories in terms of a web of 5-branes suspended between 7-branes \cite{Benini:2009gi}.
The derivation of the mirror by the brane construction guarantees that the dimensions of the Coulomb and Higgs branches of a Sicilian theory are exchanged with respect to those of the mirror theory we propose.

Since $N$ 5-branes on a torus realize 4d $\cN=4$ $\U(N)$ Super Yang-Mills (SYM),
the web construction suggests that it is possible to realize a 3d Sicilian theory by putting $\cN=4$ $\U(N)$ SYM on a graph, made of segments and trivalent junctions.
We can thus phrase our observation on mirror symmetry in terms of the S-duality of boundary conditions of $\cN=4$ SYM, following \cite{Gaiotto:2008sa,Gaiotto:2008ak}.
Our statement is then that the boundary condition of $\cN=4$ $\U(N)^3$ SYM which breaks $\U(N)^3$ down to the diagonal $\U(N)$ subgroup is S-dual to the boundary condition that breaks the group to $\U(N)^3/\U(1)_\text{diag}$ and couples it to $T_N$.

We can in fact work in a purely field theoretic framework, considering $\cN=4$ SYM on a graph. Firstly, for all 3d theories that can be realized in this way, mirror symmetry is a ``modular'' operation applied to each constituent separately. Secondly, this perspective allows to enlarge the class of theories beyond string theory constructions.

Formulating the problem in terms of $\cN=4$ SYM on a graph is particularly useful to find the mirror of Sicilian theories of type $D_N$ \cite{Tachikawa:2009rb}, obtained by compactification of the 6d $\cN=(2,0)$ $D_N$ theory or equivalently $N$ M5-branes on top of an M-theory orientifold. In this case we do not have a IIB brane construction of the junction. Nevertheless the framework of $\cN=4$ SYM on a graph will give us the mirror.

The structure of the paper is as follows. We set some conventions in section~\ref{sec:conventions}. In section~\ref{sec:brane construction} we use the 5-brane web construction to find the mirror of the $T_N$ theory, which is the basic building block of Sicilian theories, while in section~\ref{sec:mirror sicilian} we extend the map to generic Sicilian theories.  After quickly reviewing the discussion of \cite{Gaiotto:2008ak} about half-BPS boundary conditions, in section~\ref{sec:boundary conditions} we rephrase our mirror map in terms of $\cN=4$ SYM on a graph. This language is exploited in section~\ref{sec:DN Sicilian} to find the mirror of $D_N$ Sicilian theories.
We conclude in section~\ref{sec:discussion} providing many open directions.
In appendix \ref{app:Hitchin}, we explicitly check that the Coulomb branch of Sicilian theories is equal to the Higgs branch of their mirror quivers, by studying the moduli space of the Hitchin system. In appendix~\ref{app:rhovee} we discuss some properties of the S-dual of D-type punctures.

\section{Rudiments of 3d $\cN=4$ theories}
\label{sec:conventions}

3d $\cN=4$ theories have a constrained moduli space \cite{Intriligator:1996ex}: it can have a Coulomb branch, parameterized by massless vector multiplets, and a Higgs branch, parameterized by massless hypermultiplets.  There can be mixed branches as well, parameterized by both sets of fields. All branches are hyperk\"ahler.
When the theory is superconformal, it has R-symmetry $\SO(4)_R\simeq \SO(3)_X \times \SO(3)_Y$: then $\SO(3)_X$ acts on the lowest component of vector multiplets, while $\SO(3)_Y$ acts on that of hypermultiplets.

Both the Coulomb and Higgs branch can support the action of a global non-R symmetry group:
we will call them Coulomb and Higgs symmetries respectively. When the 3d theory has a Lagrangian description, the Higgs branch is not quantum corrected and the Higgs symmetry is easily identified as the action on hypermultiplets.
Coulomb symmetries are  subtler. The classic example is a $\U(1)$ vector multiplet with field strength $F$: then $J = *F=d\phi$ is the conserved current of a $\U(1)$ Coulomb symmetry, which shifts the dual photon $\phi$. Quantum corrections can enhance the Abelian Coulomb symmetry to a non-Abelian one.

To both Coulomb and Higgs symmetries are associated conserved currents. We will often use them to ``gauge together'' two or more theories. What we  mean are the following two options. Firstly, we can take two theories---each of which has a global symmetry group $G$ acting on the Higgs branch---then take the current of the diagonal subgroup and couple it to a $G$ vector multiplet, in a manner which is  $\cN=4$ and gauge invariant.
Secondly, we can take two theories---each of which has a global symmetry group $G$ acting on the Coulomb branch---and couple a $G$ vector field to the diagonal subgroup. To preserve $\cN=4$ supersymmetry,
one needs to use a \emph{twisted} vector multiplet whose lowest component is non-trivially acted by $\SO(3)_Y$.
Twisted vector multiplets can also be coupled to
twisted hypermultiplets, whose lowest component is non-trivially acted by $\SO(3)_X$.
The mirror map then relates two theories $\cA$ and $\cB$, such that the Coulomb branch of $\cB$ is the Higgs branch of $\cA$ and vice versa.

We will often consider 5d, 4d and 3d versions of a theory. What we mean is that a lower dimensional version is obtained by simple compactification on $S^1$. When a 4d $\cN=2$ theory is compactified to 3d there is a close relation between the moduli spaces of the two versions \cite{Seiberg:1996nz}. The Higgs branches are identical.
If the 4d Coulomb branch has complex dimension $n$, the 3d Coulomb branch is a fibration of $T^{2n}$ on the 4d Coulomb branch, and has quaternionic dimension $n$.
The K\"ahler class of the torus fiber is inversely proportional to the radius of $S^1$.
We often take the small radius limit and discuss the resulting superconformal theory.

\section{Mirror of triskelions via a brane construction}
\label{sec:brane construction}

The objective of this section is to find the mirror of the $T_N$ theory, and more generally of triskelion theories.
In the next section, we will explain how to gauge them together and construct the mirror of general Sicilian theories.

\subsection{Mirror of $T_N$}
\label{sec:mirror of TN}

The mirrors of a large class of theories have been found
by Hanany and Witten by exploiting a brane construction \cite{Hanany:1996ie} (see also \cite{deBoer:1996mp,deBoer:1996ck}): one realizes the field theory as the low energy limit of a system in IIB string theory of D3-branes suspended between NS5-branes and D5-branes.
The mirror theory is obtained by performing an S-duality on the configuration, and then reading off the new gauge theory.
We cannot apply this program directly to the M-theory brane construction of Sicilian theories, except for those cases that reduce to a IIA brane construction.

A 3d theory can also be studied by first constructing its 5d version using a web of 5-branes and then compactifying it on $T^2$ \cite{Aharony:1997ju}.
In \cite{Benini:2009gi} it was shown how to lift the Sicilian theories to five dimensions, and how to get a brane construction of them in IIB string theory. That paper focused on the uplift of $N$ M5-branes wrapped on the sphere with three generic punctures, and this is all we need to start.

\begin{figure}[t]
\begin{center}
\begin{minipage}{.7\textwidth}
\begin{tabular}{|r|ccccc|cc|ccc|}
\hline
 & 0 & 1 & 2 & 3 & 4 & 5 & 6 & 7 & 8 & 9 \\
\hline
\hline
D5 & $-$ & $-$ & $-$ & $-$ & $-$ & $-$ & & & & \\
NS5 & $-$ & $-$ & $-$ & $-$ & $-$ & & $-$ & & & \\
(1,1) 5-brane & $-$ & $-$ & $-$ & $-$ & $-$ & \multicolumn{2}{c|}{\text{angle}} & & & \\
(p,q) 7-brane & $-$ & $-$ & $-$ & $-$ & $-$ & & & $-$ & $-$ & $-$ \\
\hline
\end{tabular}
\end{minipage} \hspace{\stretch{1}} \begin{minipage}{.26\textwidth}
\includegraphics[width=\textwidth]{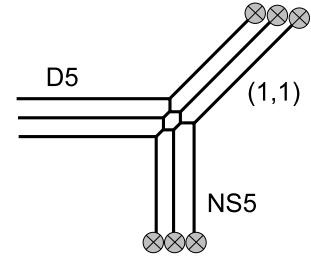}
\end{minipage}

\caption{Left: Table of directions spanned by the objects forming the web. Right: The web of $N$ D5-branes, NS5-branes and $(1,1)$ 5-branes; here $N=3$. In the figure the D5's are semi-infinite, while NS5's and $(1,1)$ 5-branes terminate each on a 7-brane $\otimes$ of the same type. The Coulomb branch of the 5d low energy theory is not sensitive to this difference.
\label{fig:p,q web}}
\end{center}
\end{figure}

Consider a web of semi-infinite 5-branes in IIB string theory, made of $N$ D5-branes, $N$ NS5-branes and $N$ $(1,1)$ 5-branes meeting at a point, as summarized in figure \ref{fig:p,q web}. At the intersection lives a 5d theory which we call the 5d $T_N$ theory \cite{Benini:2009gi},
and many properties of its Coulomb branch can be read off the brane construction.
Instead of keeping the 5-branes semi-infinite, we can terminate each of them at finite distance on a 7-brane of the same $(p,q)$-type as in figure \ref{fig:p,q web}.
The distance does not affect the Coulomb branch of the low energy 5d theory: a $(p,q)$ 5-brane terminating on a $(p,q)$ 7-brane on one side and on the web on the other side has boundary conditions that kill all massless modes \cite{Hanany:1996ie}. However this modification is useful for three reasons:
it displays the Higgs branch as normalizable deformations of the web along $x^{7,8,9}$; it admits a generalization where multiple 5-branes end on the same 7-brane (this configurations are related to generic punctures on the M5-branes, as in section~\ref{sec:mirror of triskelion}); upon further compactification to three dimensions, it allows us to read off the mirror theory.

Our strategy to understand the 3d $T_N$ theory is to consider the IIB brane web on $T^2$, understand the low energy field theory leaving on each of the three arms separately, and finally understand how they are coupled together at the junction. We exploit the brane construction here, and present a different perspective in section \ref{sec:boundary conditions}.

Consider, for definiteness, the arm made of $N$ D5-branes ending on $N$ D7-branes. We first want to consider the arm alone,
therefore we will substitute the web junction with a single D7-brane.
Since the brane construction lives on $T^2$, we can perform two T-dualities and one S-duality to map it to a system of D3-branes suspended between NS5-branes---the familiar Hanany-Witten setup.
We identify the $\SO(3)$ symmetry rotating $x^{7,8,9}$ with $\SO(3)_Y$.
$\SO(3)_X$
will only appear in the low-energy limit, rotating the motion in the $x^{5,6}$-plane and the Wilson lines around the torus.

\begin{figure}[t]
\begin{center}
\includegraphics[width=.5\textwidth]{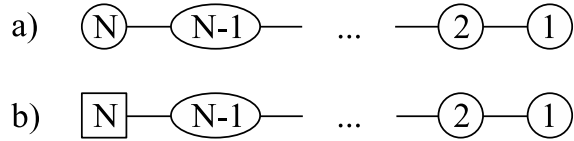}
\caption{a) Quiver diagram resulting from a configuration of D3-branes suspended between NS5-branes. b) Quiver diagram of the $T[\SU(N)]$ theory. Circles are $\U(r_a)$ gauge groups, the square is an $\SU(N)$ global symmetry group and lines are bifundamental hypermultiplets.
\label{fig:T(SU(N))}}
\end{center}
\end{figure}

The low energy field theory is a linear quiver of unitary gauge groups, as in figure \ref{fig:T(SU(N))}a. Each stack of $r$ D3-branes leads to a $\U(r)$ twisted vector multiplet,
while each NS5-brane
leads to a twisted bifundamental hypermultiplet. The other two arms made of $(p,q)$ 5-branes and 7-branes lead to the same field theory: to read it off, we perform first an S-duality to map the system to D5-branes and D7-branes, and then proceed as above.%
\footnote{The gauge couplings at intermediate energies will be different, but this will not affect the common IR fixed point to which the theories flow.}

\begin{figure}[t]
\[
\includegraphics[height=4.5cm]{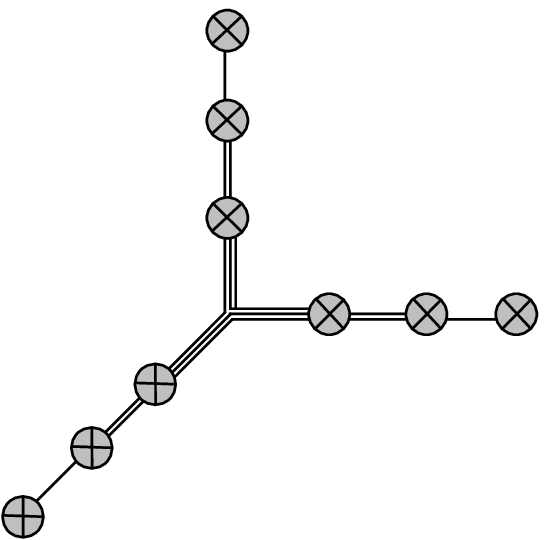}
\qquad\qquad
\includegraphics[height=4.5cm]{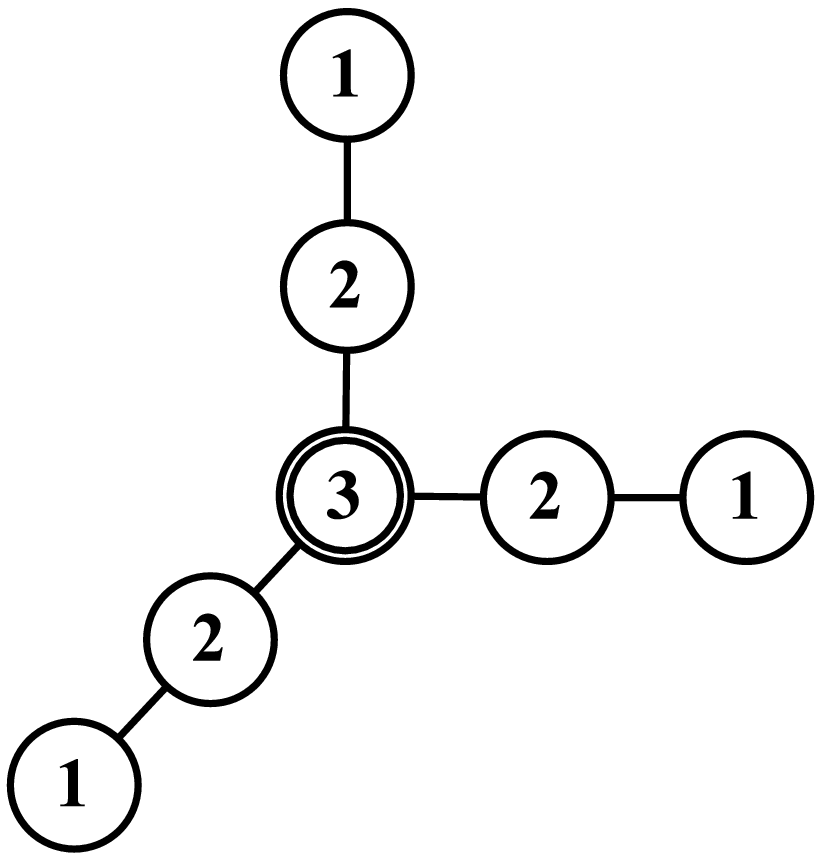}
\]
\caption{Left: $(p,q)$-web realizing the $T_N$ theory, with aligned 7-branes. Right: Quiver diagram of the mirror of $T_N$, with gauge groups $\U(r)$. The group at the center is taken to be $\SU$, to remove the decoupled overall $\U(1)$. Here $N = 3$.
\label{fig:TNmirror}}
\end{figure}

To conclude, we need to understand what is the effect of joining the three arms together, instead of separately ending each of them on a single $(p,q)$ 7-brane.
We look at the effect on the moduli space: In each arm, the motion of the 5-branes along $x^{7,8,9}$
is parameterized by the twisted vector multiplet. When the three arms are joined together, the positions of the 5-branes at the intersection are forced to be equal, therefore the boundary condition breaks the three $\U(N)$ gauge groups to the diagonal one. The resulting low energy field theory is a quiver gauge theory, depicted in figure \ref{TN} and \ref{fig:TNmirror}, that we will call \emph{star-shaped}.
Notice that the $\U(1)$ diagonal to the whole quiver is decoupled; this can be conveniently implemented by taking the gauge group at the center to be $\SU(N)$.
Interestingly, we find that the mirror theory of  $T_N$ has a simple Lagrangian description.

In view of the subsequent generalizations, it is useful to give a slightly different but equivalent definition of the star-shaped quiver: To each maximal puncture we associate a 3d linear quiver, introduced in \cite{Gaiotto:2008ak} and called $T[\SU(N)]$.%
\footnote{Note that this theory is distinct from the $T_N$ theory.}
Its gauge group has the structure
\begin{equation}
\underline{\SU(N)} - \U(N-1) - \U(N-2) - \cdots - \U(1) \;,
\end{equation} see figure \ref{fig:T(SU(N))}b. The underlined group is a flavor symmetry, and we have bifundamental hypermultiplets between two groups.
The $\SU(N)$ Higgs symmetry is manifest,
whilst only the Cartan subgroup of the $\SU(N)$ Coulomb symmetry is manifest and enhancement is due to monopole operators.
The star-shaped quiver is then obtained by taking three $T[\SU(N)]$ quivers, one for each arm, and gauging together the three $\SU(N)$ Higgs symmetries.

\subsection{Mirror of triskelion}
\label{sec:mirror of triskelion}

We can generalize the mirror symmetry map to 3d triskelion theories. 4d triskelion theories are the low energy limit of $N$ M5-branes wrapped on the Riemann sphere with three generic punctures.
A class of half-BPS punctures is classified by Young diagrams with $N$ boxes \cite{Gaiotto:2009we}: we will indicate them as $\rho=\{h_1, \dots, h_J\}$ where $h_1 \geq \dots \geq h_J$ are the heights of the columns, and $J$ is the number of columns.

Such classification arises naturally in the IIB brane construction \cite{Benini:2009gi}:
we allow multiple 5-branes to terminate on the same 7-brane. For each arm, the possible configurations are labeled by partitions of $N$, that is Young diagrams $\rho=\{h_1,\dots,h_J\}$. In our conventions, $J$ is the number of 7-branes and $h_a$ is the number of 5-branes ending on the $a$-th 7-brane. The maximal puncture considered before is $\{1,\dots,1\}$.
The global symmetry at each arm is easily read off as
\be
\label{symmetry of puncture}
G_\rho = \text{S} \Big( \prod\nolimits_h \U(N_h) \Big) \;,
\ee
where $N_h$ is the number of columns of $\rho$ of height $h$, and the diagonal $\U(1)$ has been removed. The brane construction also makes clear that a triskelion theory with punctures $(\rho_1,\rho_2,\rho_3)$ arises as the effective theory along the Higgs branch of $T_N$: it can be obtained by removing 5-branes suspended between 7-branes, and this is achieved by moving along the Higgs branch.

\begin{figure}[t]
\begin{center}
\includegraphics[width=.9\textwidth]{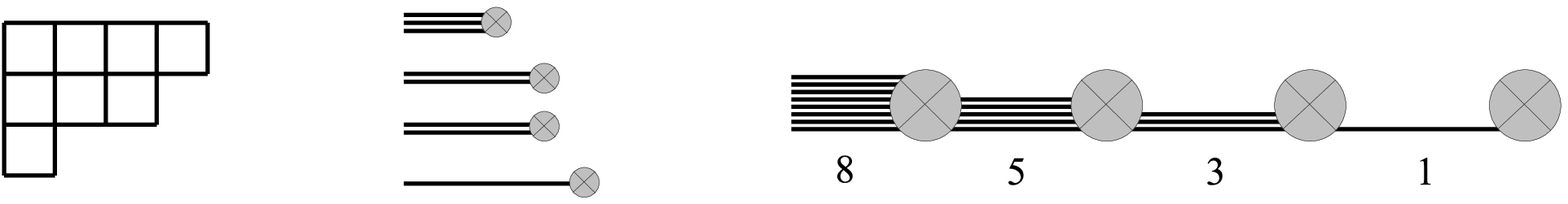}
\caption{Mirror theory of a generic puncture. Left: Young diagram of the puncture $\{3,2,2,1\}$. Its global symmetry is $\U(2) \times \U(1)$. Center: Corresponding configuration of 5-branes and 7-branes $\otimes$. Right: The same configuration, with the 7-branes aligned. The ranks are then read off to be 8, 5, 3, 1.
\label{fig:generic puncture}}

\bigskip

\bigskip

\includegraphics[width=.3\textwidth]{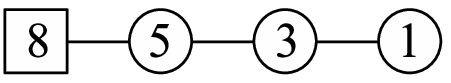}
\caption{$T_\rho[\SU(N)]$ theory, for $\rho = \{3,2,2,1\}$. All gauge groups are $\U(r)$.
\label{fig:eg Trho}}
\end{center}
\end{figure}

To construct the mirror of the 3d triskelion theory we proceed as before. We consider the three arms separately, substituting the junction with a single 7-brane. We perform an S-T$^2$-S duality on each arm, to map it to a system of D3-branes suspended between NS5-branes; the field theory is read off to be a 3d linear quiver with unitary gauge groups.
These steps are summarized in figure \ref{fig:generic puncture}. Finally we glue together the three arms, which corresponds to setting boundary conditions that break the three $\U(N)$ factors to the diagonal $\U(N)$; the overall $\U(1)$ is decoupled and removed, thus making the gauge group at the center to be $\SU(N)$.

As before, we can construct the 3d star-shaped quiver in an equivalent way. To each puncture%
\footnote{We indicate both a Young diagram and the corresponding puncture with the same symbol $\rho$.}
$\rho=\{h_1,\dots,h_J\}$ we associate a linear quiver $T_\rho[\SU(N)]$ \cite{Gaiotto:2008ak}: it has the structure
\be
\underline{\SU(r_0)} - \U(r_1) - \U(r_2) - \ldots - \U(r_{J-1}) \;,
\ee
where the underlined group is a flavor Higgs symmetry and the others are gauge groups.
We have hypermultiplets in the bifundamental representation of $\U(r_i)\times \U(r_{i+1})$.
Here $r_a$ is given by
\be
r_a = \sum_{b=a+1}^J h_b \;.
\ee
This quiver has $\SU(N)$ symmetry on the Higgs branch and (\ref{symmetry of puncture}) on the Coulomb branch.
An example is in figure \ref{fig:eg Trho}.
The theory $T[\SU(N)]$ introduced in section \ref{sec:mirror of TN} is $T_\rho[\SU(N)]$ with $\rho=\{1,1,\ldots,1\}$, \ie{} the maximal puncture.
The 3d star-shaped quiver can be obtained by gauging together the three $\SU(N)$ Higgs symmetries of $T_{\rho_i}[\SU(N)]$ for $i=1,2,3$.

Before continuing, let us recall the structure of the Higgs and Coulomb branches of this theory \cite{Gaiotto:2008ak}. To a Young diagram $\rho=\{h_1,\ldots,h_J\}$, one associates a representation $\rho$ of $\SU(2)$ given by
\begin{equation}
\rho = \underline{h_1} \oplus \underline{h_2} \oplus \cdots \oplus \underline{h_J}\label{eq:decomp}
\end{equation}
where $\underline{h_a}$ is the irreducible $h_a$-dimensional representation.
Let the generators of $\SU(2)$ be $t^{\pm}$  and $t^3$. Then $\rho(t^+)\in \su(N)_\bC$ is the direct sum of the Jordan blocks of size $h_1$, \ldots, $h_J$. In particular this is nilpotent.
The nilpotent orbit of type $\rho$ is defined to be
\begin{equation}
\cN_\rho = \SU(N)_\bC \cdot \rho(t^+) \;.
\end{equation}  In particular it has an isometry $\SU(N)$.
Its closure $\overline{\cN_\rho}$ is a hyperk\"ahler cone and it coincides with the Higgs branch of the quiver $T_{\rho^\trans}[\SU(N)]$, where $\rho^\trans$ denotes the transpose of the Young diagram $\rho$ in which $h_a$ are the length of the rows.

The Slodowy slice $\cS_\rho$ is a certain nice transverse slice to $\cN_\rho\subset \su(N)_\bC$ at $\rho(t^+)$. The Coulomb branch of $T_\rho[\SU(N)]$ is $\cS_\rho \cap \overline\cN$, where $\cN=\cN_{\{1,\dots,1\}}$ is the maximal nilpotent orbit.  Then the isometry of the Coulomb branch is the commutant of $\rho(\SU(2))$ inside $\SU(N)$, and agrees with the symmetry \eqref{symmetry of puncture} read off from the brane construction.

\section{Mirror of Sicilian theories}
\label{sec:mirror sicilian}

After having understood the mirror of triskelions, which are the building blocks, we can proceed to generic 3d Sicilian theories.
The mirror of a 3d triskelion with punctures $(\rho_1,\rho_2,\rho_3)$ is obtained by taking the three $T_{\rho_i}[\SU(N)]$ linear quivers for $i=1,2,3$ and gauging together the three $\SU(N)$ Higgs symmetry factors.
To construct a Sicilian theory we gauge together two $\SU(N)$ Higgs symmetries, therefore on the mirror side we gauge together two $\SU(N)$ Coulomb symmetries.
In the following we study the effect of such gauging on the mirror.

\subsection{Genus zero: star-shaped quivers}

Let us consider, for simplicity, two triskelions
glued together.
The mirror is obtained by taking the two sets of linear quivers $T_{\rho_i}[\SU(N))]$ and $T_{\rho'_{i}}[\SU(N)]$, $i=1,2,3$.
We gauge together the three $\SU(N)$ Higgs symmetries in each set.
We let $\rho_1$ and $\rho_1'$ be maximal, and
gauge together the $\SU(N)$ Coulomb symmetries of $T_{\rho_1}[\SU(N)]$ and $T_{\rho'_1}[\SU(N)]$.

Since the order of gauging does not matter, we shall first consider the effect of gauging together two copies of $T[\SU(N))]$  by the $\SU(N)$ Coulomb symmetries. The resulting low energy theory \cite{Gaiotto:2008ak} has a Higgs branch $T^*\SU(N)_\bC$,
the total space of the cotangent bundle to the complexified $\SU(N)$ group, and no Coulomb branch.
The Higgs branch is acted upon by $\SU(N) \times \SU(N)$ on the left and right respectively, but every point of the zero-section breaks it to the diagonal $\SU(N)$, and no other point on the moduli space preserves more symmetry.
Since the Higgs branch has a scale given by the volume of the base space $\SU(N)_\bC$
and it is smooth, around each point the theory flows to $N^2-1$ free twisted hypermultiplets, which are then eaten by the Higgs mechanism.

\begin{figure}[t]
\noindent
\includegraphics[width=.58\textwidth]{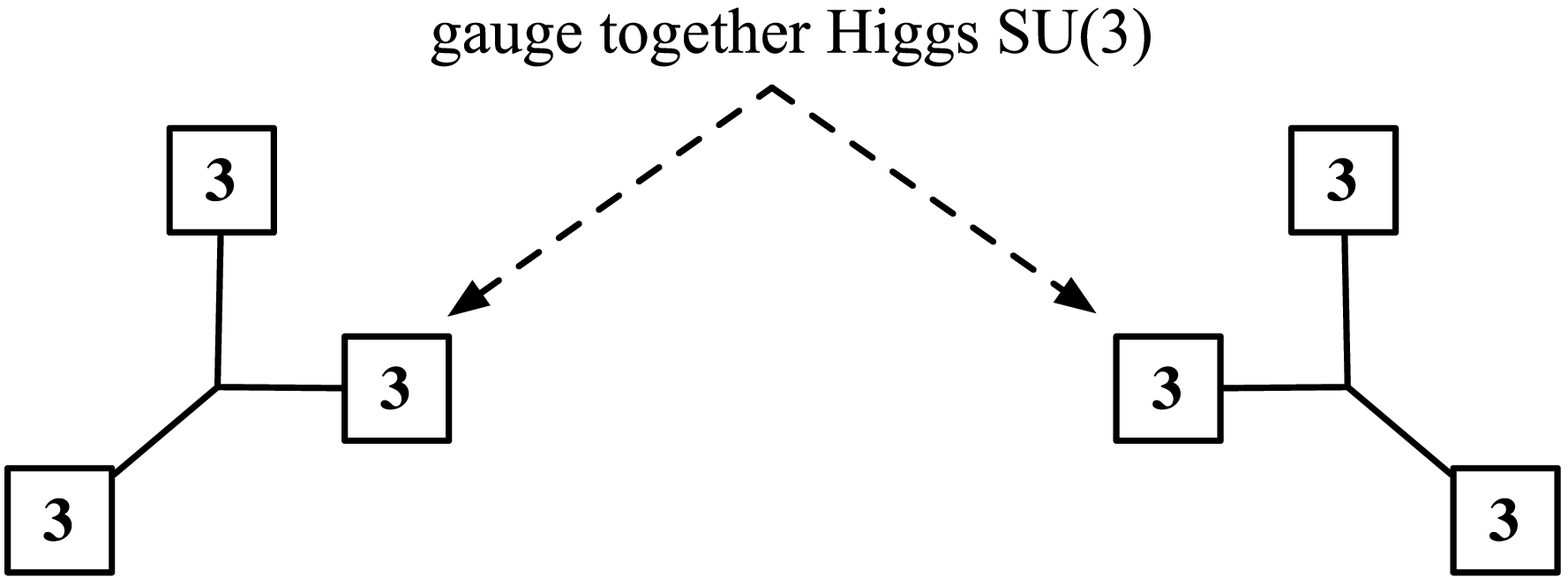}
\hspace{\stretch{1}}
\raisebox{8ex}{\Huge $\Rightarrow$}
\hspace{\stretch{1}}
\includegraphics[width=.32\textwidth]{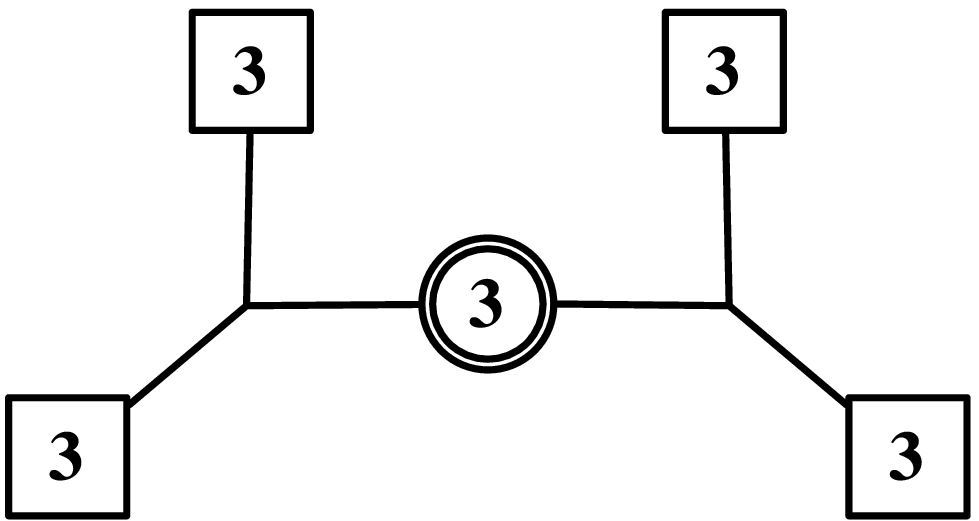}

\

\noindent
\includegraphics[width=.58\textwidth]{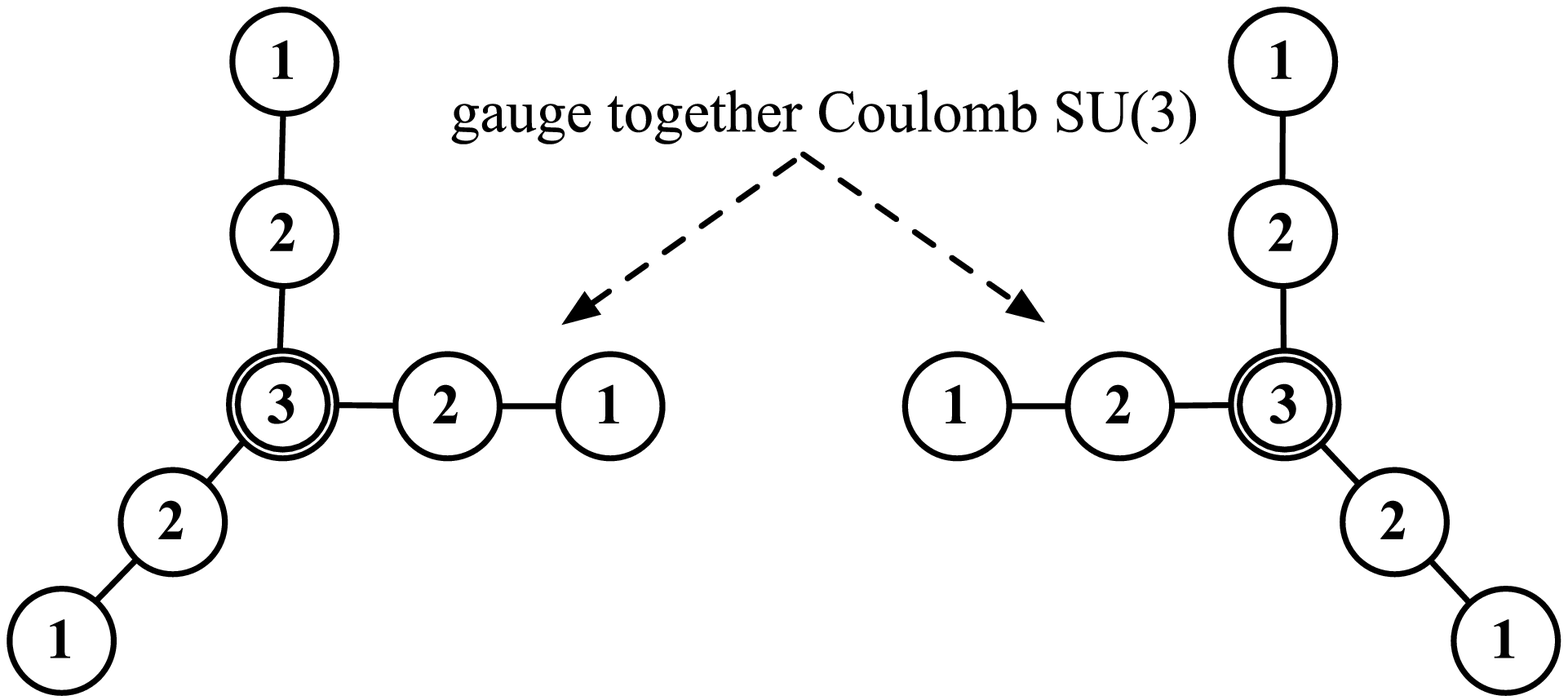}
\hspace{\stretch{1}}
\raisebox{10ex}{\Huge $\Rightarrow$}
\hspace{\stretch{4}}
\includegraphics[width=.25\textwidth]{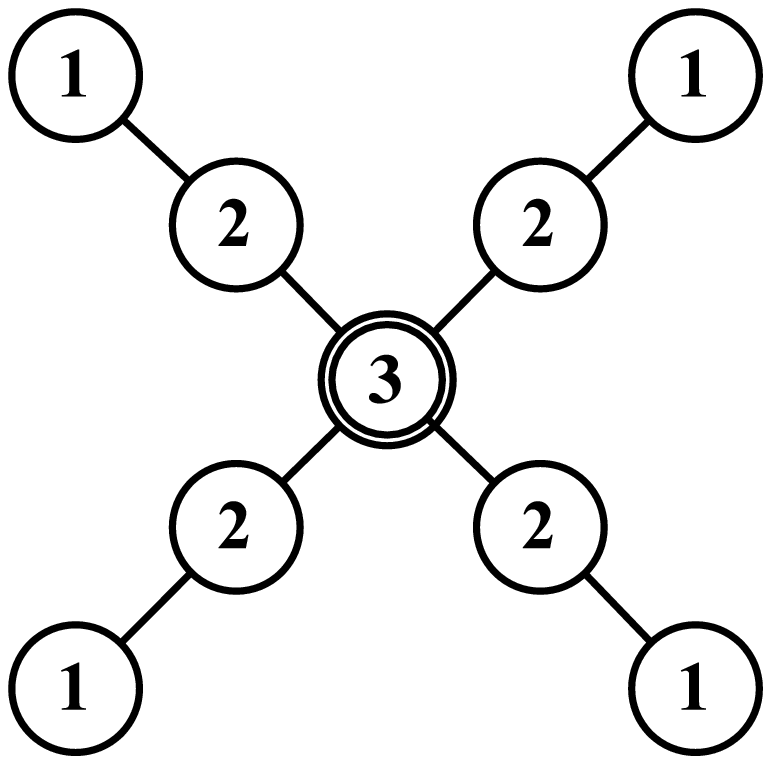}
\hspace{\stretch{3}}

\caption{Top: We take two copies of $T_N$ and gauge together two $\SU(N)$ Higgs symmetries.  Bottom: Its mirror. We gauge together two $\SU(N)$ Coulomb symmetries. This ends up eliminating the two $T[\SU(N)]$ tails. Here $N=3$.
\label{fig:gluing}}
\end{figure}

\begin{figure}[t]
\begin{center}
\hspace{\stretch{1}}
\raisebox{.5cm}{\includegraphics[width=.3\textwidth]{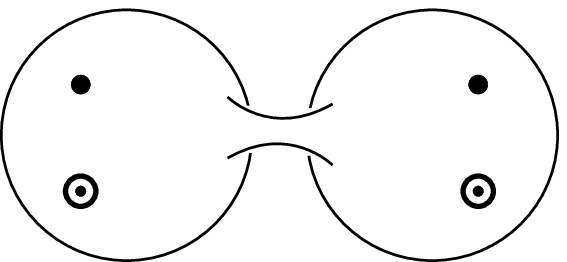}}
\hspace{\stretch{2}}
\includegraphics[width=.23\textwidth]{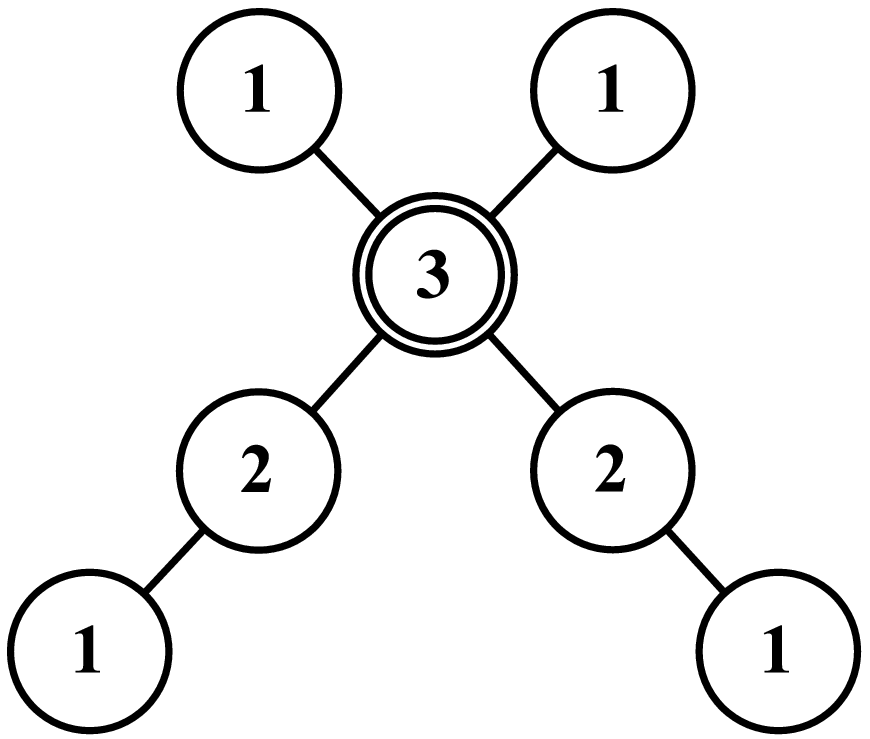}
\hspace{\stretch{1}}

\caption{Left: A pants decomposition of the Riemann surface of a genus zero Sicilian theory. In the example the four punctures are $\{1,1,1\}$ ($\odot$) and $\{2,1\}$ ($\bullet$), and the Sicilian theory is $\SU(3)$ SQCD with 6 flavors. Right: Star-shaped quiver, mirror of a genus zero Sicilian theory. Notice that the $\SU(6)\times \U(1)$ Coulomb symmetry due to monopole operators is easy to see.
\label{fig:mirror of SU3}}
\end{center}
\end{figure}

Summarizing, coupling two copies of $T[\SU(N)]$
by their $\SU(N)$ Coulomb symmetries spontaneously breaks the $\SU(N) \times \SU(N)$ Higgs symmetry to the diagonal subgroup. Therefore,  we are left with $T_{\rho_{2,3}}[\SU(N)]$ and $T_{\rho'_{2,3}}[\SU(N)]$ with all four $\SU(N)$ Higgs symmetries gauged together. See figures \ref{fig:gluing} and \ref{fig:mirror of SU3} for examples; there, the $T_N$ theory is depicted by a trivalent vertex with three boxes, each representing an $\SU(N)$ Higgs symmetry.

This is easily generalized to a generic 3d genus zero Sicilian theory obtained from a sphere with punctures.
Its mirror is obtained by taking the set of $T_\rho[\SU(N)]$ linear quivers corresponding to all punctures, and gauging all the $\SU(N)$ Higgs symmetries together.
Such theory is a star-shaped quiver.

We find that in 3d, the low energy theory only depends on the topology of the punctured Riemann surface, and not on its complex structure. This is as expected: In 4d the complex structure controls the complexified gauge couplings of the IR fixed point. When compactifying to 3d, all gauge couplings flow to infinity based on dimensional analysis, washing out the information contained therein.

\subsection{Higher genus: adjoint hypermultiplets}

Let us next consider the mirror of 3d Sicilian theories obtained from Riemann surfaces of genus $g \geq 1$.
Taking advantage of S-duality in 4d Sicilian theories, without loss of generality we can consider a pants decomposition in which all handles come from gluing together two maximal punctures on the same triskelion.

\begin{figure}[t]
\begin{center}
\includegraphics[width=.95\textwidth]{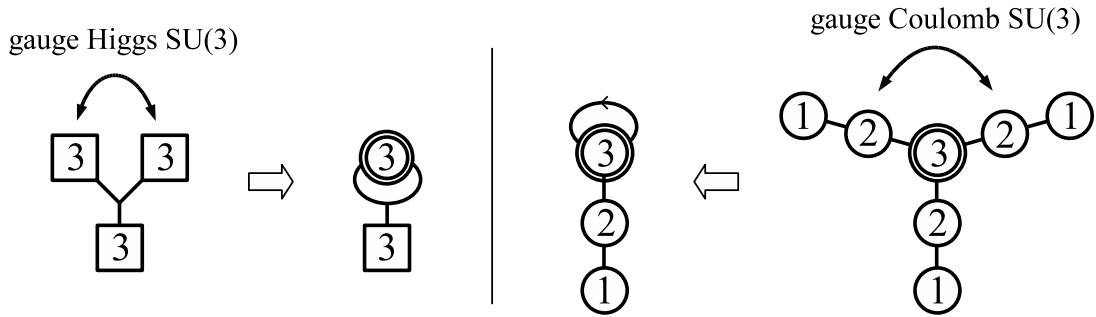}
\caption{Mirror symmetry of higher genus Sicilian theories. Left: We gauge two $\SU(N)$ Higgs symmetries producing a ``handle'' in a higher genus Sicilian theory. Right: In the mirror we gauge two $\SU(N)$ Coulomb symmetries, getting rid of two $T[\SU(N)]$ tails but leaving one adjoint hypermultiplet. Here $N=3$.
\label{fig:mirror genus}}
\end{center}
\end{figure}

The mirror can be constructed as before, by taking $T_\rho[\SU(N)]$ for each of the punctures, and suitably gauging together the Higgs and Coulomb $\SU(N)$ symmetries.
The only difference compared to the genus zero case is
that, for each of the $g$ handles, we get two copies of $T[\SU(N)]$ gauged together both on the Higgs and Coulomb branch.
This amounts to gauging the diagonal subgroup of the $\SU(N) \times \SU(N)$ Higgs symmetry of $T^*\SU(N)_\bC$, which is not broken along the zero-section: the $N^2-1$ twisted hypermultiplets living there are thus left massless.
They transform in the adjoint representation of the diagonal subgroup. See figure \ref{fig:mirror genus} for an example.

\begin{figure}[t]
\begin{center}
\includegraphics[width=.9\textwidth]{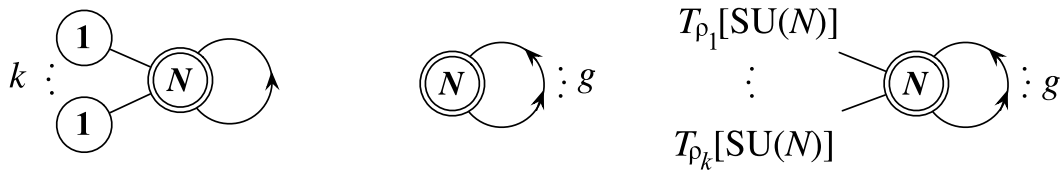}
\caption{Star-shaped mirrors of genus $g$ Sicilian theories. Left: Mirror of the Sicilian theory of genus 1 with $k$ simple punctures $\{N-1,1\}$. The Sicilian theory is a closed chain quiver (also called elliptic quiver) of $k$ $\SU(N)$ gauge groups and bifundamentals. Center: Mirror of the Sicilian theory of genus $g$ and without punctures, which is the 3d compactification of the field theory dual to the Maldacena-Nu\~nez supergravity solution. Right: Mirror of a generic genus $g$ Sicilian theory with $k$ punctures $\rho_1$, \dots, $\rho_k$. The $g$ hypermultiplets are in the adjoint of $\SU(N)$, and a $\USp(2g)$ IR symmetry emerges.
\label{fig:mirror of MN}}
\end{center}
\end{figure}

We find that the mirror theory is, as before, a gauge theory.
It is obtained by taking the set of $T_\rho[SU(N)]$ linear quivers corresponding to the punctures, plus $g$ twisted hypermultiplets in the adjoint representation of $\SU(N)$, and gauging all the $\SU(N)$ Higgs symmetries together. See figure \ref{fig:mirror of MN} for two examples and the generic case. The $g$ adjoint hypermultiplets carry an accidental IR $\USp(2g)$ Higgs symmetry, not present in the 4d theory.
Again, the 3d IR fixed point only depends on the topology of the defining Riemann surface.

One can check that the dimensions of the Coulomb and Higgs branch in the 3d Sicilian theories and star-shaped quivers agree, after exchange. In appendix \ref{app:Hitchin} we provide a half proof of the mirror symmetry map: we explicitly show that the Coulomb branch of Sicilian theories coincides with the Higgs branch of star-shaped quivers.

Let us stress two nice examples of mirror symmetry. One is the 4d theory dual to the Maldacena-Nu\~nez supergravity solution \cite{Maldacena:2000mw} of genus $g$. This theory in non-Lagrangian, however after compactification to three dimensions its mirror is $\SU(N)$ with $g$ adjoint hypermultiplets (center in figure~\ref{fig:mirror of MN}). The other example is the rank-$k$ $E_{6,7,8}$ theories. Their 3d mirror is a quiver of groups $\U(k\,n_i)$ where the shape is the extended Dynkin diagram $\hat E_{6,7,8}$ and the ranks are $k$ times the Dynkin index $n_i$ of the $i$-th node. For $k=1$ it is the example considered in the seminal paper \cite{Intriligator:1996ex}.

\section[Boundary conditions, mirror symmetry and $\cN=4$ SYM on a graph]{Boundary conditions, mirror symmetry \\ and $\cN=4$ SYM on a graph}
\label{sec:boundary conditions}

In the last section we described how to obtain the mirror of 3d Sicilian theories in terms of junctions of 5-branes compactified on $T^2$.
Since a stack of $N$ 5-branes compactified on $T^2$ gives $\cN=4$ super Yang-Mills, it is possible to rephrase what we derived from the brane construction in terms of half-BPS boundary conditions of $\cN=4$ super Yang-Mills, as was in \cite{Gaiotto:2008ak}. This perspective allows us to extend the mirror symmetry map to more general theories, not easily engineered with M5-branes.
Let us start by  reviewing the framework of \cite{Gaiotto:2008sa,Gaiotto:2008ak}.

\subsection{Half-BPS boundary conditions: review}
\label{BPS}

Consider $\cN=4$ super Yang-Mills with gauge group $G=G_1\times G_2\times \cdots$ on a half-space $x^3>0$.
In the following we set all $\theta$ angles to zero.
We introduce the metric on the Lie algebra
$ \fg=\fg_1\oplus \fg_2 \oplus \cdots$
using the coupling constants, as in
\begin{equation}
\langle a,b\rangle_{\fg}= g_1^{-2} \langle a_1,b_1\rangle_{\fg_1} +
g_2^{-2} \langle a_2,b_2\rangle_{\fg_2} + \cdots \label{KillingMetric}
\end{equation}
where
$a=a_1\oplus a_2 \oplus \cdots$ and
$b=b_1\oplus b_2 \oplus \cdots$
are two elements of $\fg$,  $\langle\cdots\rangle_{\fg_i}$ is the standard Killing metric on $\fg_i$, and $g_i$ is the coupling constant of the $i$-th factor.
The Lagrangian is then given by  \begin{equation}
S=\int d^4x \langle F_{\mu\nu},F^{\mu\nu} \rangle +
\langle D_\mu \Phi_i, D^\mu \Phi^i\rangle + \text{fermions} \;.
\end{equation}
We split the six adjoint scalar fields $\Phi_{1,\dots,6}$ into $\vec X=(X_1,X_2,X_3)$ and
$\vec Y=(Y_1,Y_2,Y_3)$.
Out of the $\SU(4)$ R-symmetry,
the symmetry manifest under this decomposition is the subgroup $\SO(3)_X\times \SO(3)_Y$, which we can identify with the $\SO(4)_R$ symmetry of a 3d $\cN=4$ CFT, as was discussed in section~\ref{sec:conventions}.

The boundary condition studied in \cite{Gaiotto:2008sa} consists of the data $(\rho,H,\cB)$.
First, $\rho$ is an embedding
\begin{equation}
\rho: \SU(2)\to G
\end{equation}
which controls the divergence of $\vec X$:
\begin{equation}
X_i \sim \frac{\rho(t_i)}{x^3} \;,
\end{equation}
where $t_i$ $(i=1,2,3)$ are three generators of $\SU(2)$.
The gauge field close to $x^3=0$ needs to commute with $\rho(\SU(2))\subset G$.
Therefore let $H$ be a subgroup of $G$ that commutes with $\rho(\SU(2))$, and
$\cB$ be a 3d $\cN=4$ CFT living on the boundary with $H$ global symmetry.
The theory $\cB$ can possibly be an empty theory, $\varnothing$.
The boundary conditions we impose are
\begin{subequations}
\label{boundary conditions}
\begin{align}
0&= F^+_{3a}| \;, \qquad & 0 &= F^-_{ab}| \;, \label{foo} \\
0&= \vec X^+ + \vec\mu_\cB| \;, & 0 &= D_3 \vec X^-| \;, \label{bar} \\
0&= D_3 \vec Y^+| \;, & 0 &= \vec Y^-| \;. \label{baz}
\end{align}
\end{subequations}
Here the indices $a,b=0,1,2$ are the directions along the boundary,
and the bar $|$ means the value at the boundary $x^3=0$.
We decompose the algebra of $G$ as $\fg = \fh \oplus \fh^\perp$.
Then the superscript $^+$ is the projection onto $\fh$ and
the superscript $^-$ the projection onto $\fh^\perp$.
Finally $\vec\mu_\cB$ is the moment map of the $H$ symmetry on the Higgs branch of $\cB$.
The condition \eqref{foo} means that on the boundary only the gauge field in $H$ is non-zero;
in other words the boundary condition sets Neumann boundary conditions in the
subalgebra $\fh$ and Dirichlet boundary conditions in the orthogonal complement $\fh^\perp$.
Note that this class of boundary conditions does \emph{not} treat $\vec X$ and $\vec Y$ equally: we will denote the boundary conditions more precisely as $(\rho,H,\cB)_{X,Y}$ when necessary.

We can define a boundary condition $(\rho',H',\cB')_{Y,X}$  where the role of $\vec X$ and $\vec Y$ is interchanged; in particular we will have $\vec Y^+ + \vec\mu_{\cB'}=0$, where $\vec\mu_{\cB'}$ is the moment map of $H'$ on the Higgs branch of the twisted hypermultiplets of $\cB'$.
The S-duality of $\cN=4$ SYM in the bulk $x^3>0$ acts non-trivially on spinors, and it is known to map the class of boundary conditions $(\rho,H,\cB)_{X,Y}$ to another one with the role of $\vec X$ and $\vec Y$ exchanged:
\begin{equation}
S: (\rho,H,\cB)_{X,Y} \mapsto (\rho',H',\cB')_{Y,X} \;.
\end{equation}
Let us emphasize again that this involves the exchange of the role of untwisted and twisted multiplets of the boundary 3d theory, and it is closely related to mirror symmetry.

For $G=\SU(N)$, it was shown in \cite{Gaiotto:2008ak} that
\begin{equation}
S:\;
\big( 1,\, \SU(N),\, T_\rho[\SU(N)] \big)_{X,Y}
\;\mapsto\;
\big( \rho, \, 1,\, \varnothing \big)_{Y,X}
 \;.\label{bcrho}
\end{equation}

For example, $\rho = \{1,\dots,1\}$, which we abbreviate as just $\rho=1$, is the trivial embedding and $(1,\, 1,\, \varnothing)$ is the standard Dirichlet boundary condition, which can be realized by ending $N$ D3's on $N$ D5-branes. Its S-dual has $T[\SU(N)]$ on the boundary,
and comes from ending $N$ D3's on $N$ NS5-branes.
On the other extreme, the theory $T_{\{N\}}[\SU(N)]$ is an empty theory and
$\big( 1,\, \SU(N),\, \varnothing \big)$ is the Neumann boundary condition. This can be realized by ending $N$ D3's on 1 NS5-brane (and decoupling the $\U(1)$). Its S-dual is $(\{N\},\, 1,\, \varnothing)$, and corresponds to ending $N$ D3's on $1$ D5-brane.

\begin{figure}[t]
\begin{minipage}{.55\textwidth}
\small
\[
\begin{array}{rcccc|c|cc|ccccc}
& & 0 & 1 & 2 & 3 & 4 & 5 & 6 & 7 & 8 & 9 & 10 \\
\hline
\multirow{2}{*}{IIB \Big\{\!\!} & $D3$ & - & - & - & - & & \square  \\
& $D5$ & - & - & - & & & \square & - & - & - \\
\hline
\multirow{2}{*}{IIA \Big\{\!\!} & $D4$ & - & - & - & - & - & \square \\
& $D6$ & - & - & - & & - & \square & - & - & - \\
\hline
\multirow{2}{*}{M \Big\{\!\!} & $M5$ & - & - & - & - & - & - \\
& $KK$ & - & - & - & & - & & - & - & - \\
\hline
\multirow{2}{*}{IIA \Big\{\!\!} & $D4$ & - & - & - & - & \square & - \\
& $KK$ & - & - & - & & \square & & - & - & - \\
\hline
\multirow{2}{*}{IIB \Big\{\!\!} & $D3$ & - & - & - & - & \square & \\
& $NS5$ & - & - & - & & \square & - & - & - & -
\end{array}
\]
\end{minipage}
\hspace{\stretch{1}}
$ \vcenter{\hbox{\includegraphics[scale=.5]{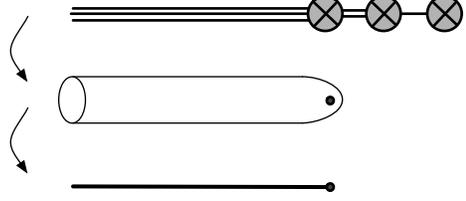}}} $

\caption{Boundary conditions of 4d $\cN=4$ SYM from 5d $\cN=2$ SYM.
By T-duality, an arm of 5-branes ending on 7-branes (top of figure) is mapped to D3/D5 or D4/D6 (first two lines of table). Uplift to M-theory gives M5-branes on a punctured cigar---a complex surface in the KK monopole---and further reduction gives D4-branes on the same cigar (middle of figure). Reduction along the $S^1$ of the cigar gives 4d $\cN=4$ SYM on a half-space $x^3 \geq 0$ with half-BPS boundary conditions (bottom of figure).
In the table we perform T-duality and uplift/reduction along $x^{4,5}$. A square means coordinates to be removed.
\label{fig:puncture-5dSYM}}
\end{figure}

This pair of boundary conditions arise naturally from 5-branes ending on 7-branes,
see figure~\ref{fig:puncture-5dSYM}.
Start from D5-branes ending on D7-branes. Compactification on $T^2$ and T-duality leads to a configuration of $N$ D3-branes ending on D5-branes.
This realizes the boundary condition $(\rho,1,\varnothing)$ of 4d $\cN=4$ SYM , on the right of (\ref{bcrho}).  When only one T-duality is performed, it can also be thought of as $N$ M5-branes on a cylinder ending on a cap with a puncture inserted, or a punctured cigar, further compactified on $S^1$.
Then it can be thought of as $N$ D4-branes on the same cigar geometry.
The Kaluza-Klein reduction along the $S^1$ of the cigar produces 4d SYM on a half-space, and since the original system preserves half of the supersymmetry, the boundary condition is also half-BPS. In fact, this corresponds to $N$ D3-branes ending on NS5-branes, and realizes the boundary condition $(1,\SU(N),T_\rho[\SU(N)])$ of 4d SYM on the left of (\ref{bcrho}): we have just performed S-duality.

\subsection{Junction and boundary conditions from the brane web}

In our brane construction, $N$ $(p,q)$ 5-branes on the torus give $\cN=4$ $\U(N)$ super Yang-Mills.
The 6d gauge coupling of a $(p,q)$ 5-brane is inversely proportional to its tension. Compactifying on $T^2$ and performing S-duality of the resulting 4d theory,
its action is given schematically by
\begin{equation}
\label{eq:4daction}
T \int d^4 x\, \Bigl(  \tr F_{\mu\nu} F^{\mu\nu} + \tr \partial_\mu  X_i\partial^\mu  X_i +\tr \partial_\mu  Y_i\partial^\mu  Y_i\Bigr) \;.
\end{equation}
Here $T$ is the tension of the 5-brane multiplied by the area of $T^2$, $Y_{1,2,3}$ is a fluctuation along $x^{7,8,9}$,
$X_1$ is the fluctuation transverse to the brane inside $x^{5,6}$ and $X_{2,3}$ come from the Wilson lines around $T^2$.

\begin{figure}[t]
\begin{center}
\includegraphics[height=3.5cm]{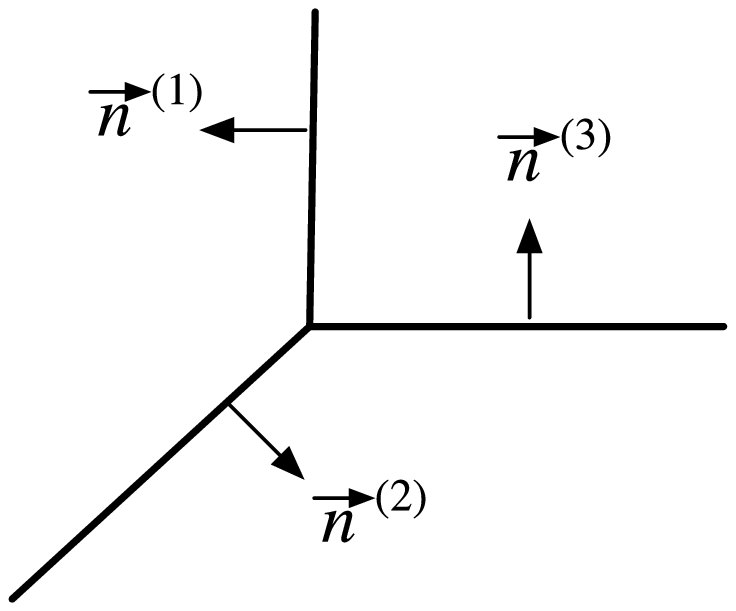}\qquad\qquad
\includegraphics[height=3.5cm]{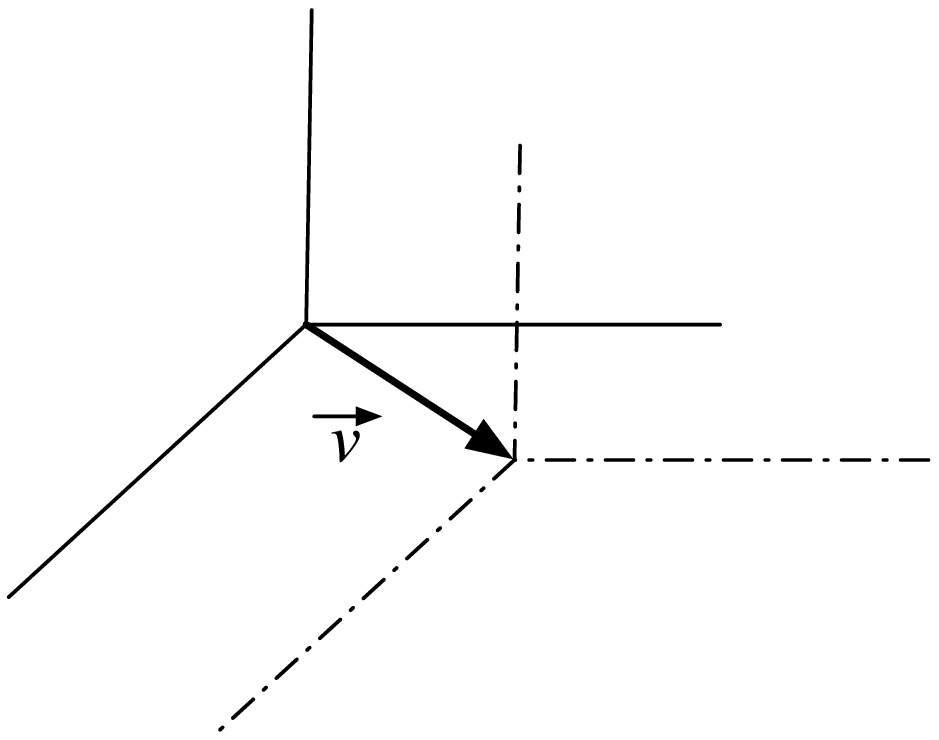}
\caption{Left: In the $x^{5,6}$-plane, we measure the worldvolume displacements $X_1^{(1,2,3)}$ of the three segments along the unit normal vectors $\vec n_{(1,2,3)}$. The $\vec Y^{(i)}$ displacements are along $x^{7,8,9}$, ``orthogonal'' to the paper.
Right: When the junction is moved by $\vec v$, the displacements are given by $X_1^{(i)}=\vec v \cdot \vec n^{(i)}.$
\label{xs}}
\end{center}
\end{figure}

We would like to understand the boundary condition corresponding to the junction of $N$ D5-, NS5- and $(1,1)$ 5-branes, see figure \ref{xs}. Let us first consider the case $N=1$.
Let us denote the unit normal to the 5-branes by $\vec n_{1,2,3}$
and the tensions of the 5-branes by $T_{1,2,3}$.
The condition of the balance of forces can be written as
\begin{equation}
T_1 \vec n_1+T_2 \vec n_2+ T_3 \vec n_3 = 0 \;.
\end{equation}
The three arms provide three copies of $\U(1)$ SYM, that we can think of as a single $\U(1)^3$ SYM.
We measure $X^{(1,2,3)}_1$ along the normal $\vec n^{(1,2,3)}$ of each of the 5-branes.
The boundary condition for the scalar $X_1$ is
$T_1 X^{(1)}_1 + T_2 X^{(2)}_1 + T_3 X^{(3)}_1=0$
that we expect to be enhanced to
\begin{equation}
\label{boo}
T_1 \vec X^{(1)} + T_2\vec X^{(2)} + T_3\vec X^{(3)}=0
\end{equation}
after compactification on $T^2$.
On the other hand, the boundary condition for  $\vec Y$ is just
\begin{equation}
\label{zot}
\vec Y^{(1)} = \vec Y^{(2)} = \vec Y^{(3)}
\end{equation}
because they can only move along $x^{7,8,9}$ together.
Comparing with the formulation in \eqref{boundary conditions},
these boundary conditions can be expressed in the two equivalent, S-dual ways
\be
\label{U(1) BC}
S: \;  (1,\, \U(1)_\text{diag},\, \varnothing)_{X,Y}
\quad\mapsto\quad
(1,\, \U(1)^3/\U(1)_\text{diag},\, \varnothing)_{Y,X} \;,
\ee
where $\U(1)_\text{diag}$ is the diagonal subgroup of $\U(1)^3$.
Note that the relation \eqref{boo} determines the orthogonal complement to the diagonal subgroup under the metric of $\U(1)^3$ \eqref{KillingMetric} given by the coupling constants.

The S-duality/mirror symmetry of the two conditions in (\ref{U(1) BC}) is easily checked. Consider the expression on the left, and close each arm at the external end with Neumann boundary conditions $(1,\U(1),\varnothing)_{X,Y}$: this gives one 3d free vector multiplet. Now consider the expression on the right and close each arm with the S-dual boundary conditions, namely Dirichlet $(1,1,\varnothing)_{Y,X}$: this gives one free twisted hypermultiplet.

\begin{figure}[t]
\[
\includegraphics[width=.9\textwidth]{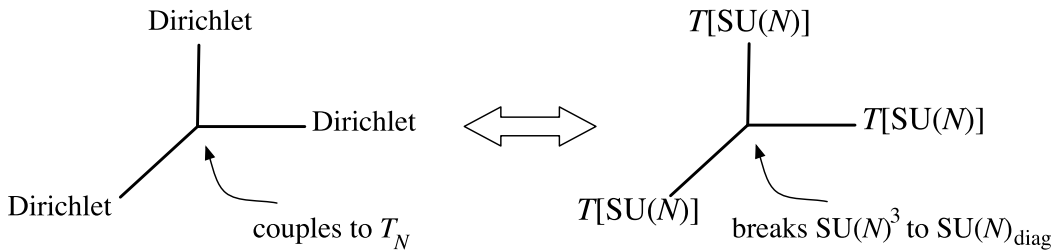}
\]
\caption{Left: $\cN=4$ $\SU(N)$ SYM on the segments, boundary conditions $(1,\SU(N)^3, T_N)$ at the center and $(1,1,\varnothing)$ at the punctures. Right: Its S-dual; boundary conditions $(1,\SU(N)_\text{diag},\varnothing)$ at the center and $(1,\SU(N),T[\SU(N)])$ at the punctures.
\label{fig:mirror BC}}
\end{figure}

For generic $N$ we still have the conditions (\ref{boo})--(\ref{zot}), as can be checked by separating the $N$ simple junctions along the $\vec Y$ direction.
The decoupled overall $\U(1)$ part is the same as before.
Then the boundary condition for the $\SU(N)$ part is
\be
\label{diagbc}
(1,\, \SU(N)_\text{diag},\, \varnothing)_{X,Y} \;.
\ee
To obtain its S-dual boundary conditions, we proceed as follows. We take a trivalent graph with $\SU(N)$ $\cN=4$ SYM on each arm, boundary conditions $\big( 1,\SU(N),T[\SU(N)] \big)$ at the external end of each arm, and the breaking-to-the-diagonal boundary condition $(1, \SU(N)_\text{diag},\varnothing)$ at the junction (see right panel in figure \ref{fig:mirror BC}). This configuration realizes, at low energy, the quiver diagram in figure~\ref{TN}.
As found in section \ref{sec:mirror of TN}, this quiver is the mirror of the $T_N$ theory.
On the other hand, we can directly perform S-duality on the configuration of SYM on a graph: on each arm we still have $\SU(N)$ SYM (which is self-dual), at the external end of each arm we get $(1, 1, \varnothing)$, while at the junction we get the boundary condition we are after (see left panel in figure \ref{fig:mirror BC}).
Since $(1,1,\varnothing)$ is the usual Dirichlet boundary condition, to obtain $T_N$ which has $\SU(N)^3$ Higgs symmetry it must be
\be
\label{S-dual BC junction}
S:\;
\big( 1,\, \SU(N)_\text{diag},\, \varnothing \big)_{X,Y}
\;\mapsto\;
\big( 1,\, \SU(N)^3,\, T_N \big)_{Y,X}
\;.
\ee

This can be proved also by considering a simple case in which we already know the mirror symmetry map.
For instance, consider the 3d Sicilian theory given by one puncture $\rho=\{N\}$ on the torus: this is 3d $\cN=8$ $\SU(N)$ SYM.
The graph construction has two $\SU(N)$ segments, $(1,\SU(N)^3,T_{N})$ at the junction and Dirichlet boundary condition at the puncture. The mirror theory is $\cN=8$ $\SU(N)$ SYM itself. The S-dual graph has $\SU(N)$ on the segments and Neumann boundary condition at the puncture. To reproduce the mirror, we need $(1,\SU(N)_\text{diag},\varnothing)$ at the junction.

\subsection{Junction and boundary conditions from 5d SYM}
\label{sec:junction}

We can derive the boundary condition of diagonal breaking (\ref{diagbc}) also from 5d SYM on the punctured Riemann surface.
The 3d $T_N$ theory arises from $N$ D4-branes on a three-punctured sphere $\cC$.
At low energy we get maximally-supersymmetric 5d SYM on $\cC$, which has $\U(N)$ gauge field $A_\mu$, curvature $F_{\mu\nu}$ and scalar fields $X_{1,2}$, $Y_{1,2,3}$.
To preserve supersymmetry, the theory
is twisted so that $X_{1,2}$ are effectively one-forms on $\cC$.
The action of the bosonic sector is roughly given by
\begin{equation}
\frac1{g^2_{5d}}\int d^5x \left[\tr F_{\mu\nu}F^{\mu\nu} + \tr D_{[\mu} X_{\nu],a}D^{[\mu} X_{a}^{\nu]} + \tr D_\mu Y_a D^\mu Y_a \right]\label{eq:5daction}
\end{equation}
where $D$ is the covariant derivative.

\begin{figure}
\[
\vcenter{\hbox{\includegraphics[scale=.45]{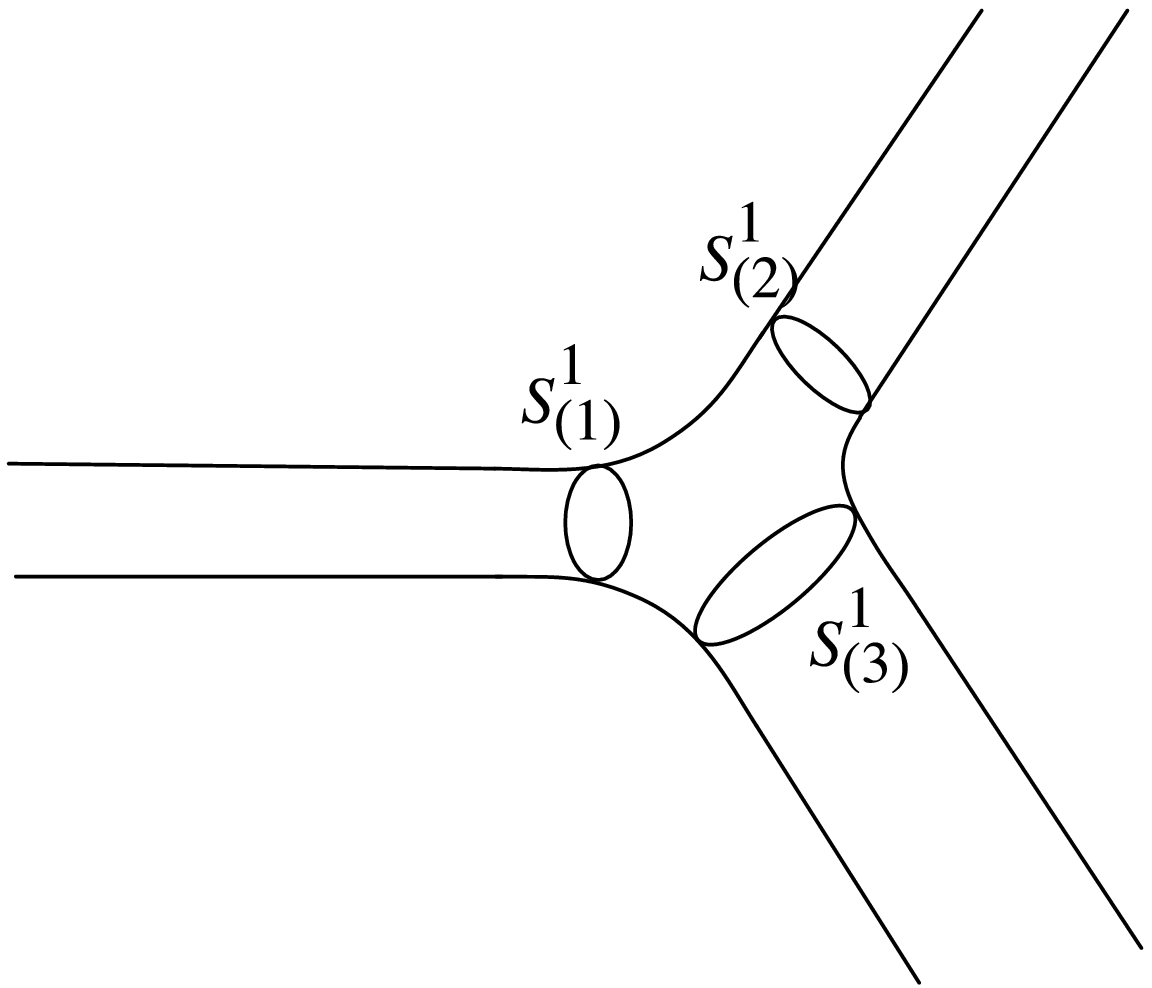}}} \qquad\Longrightarrow\qquad
\vcenter{\hbox{\includegraphics[scale=.45]{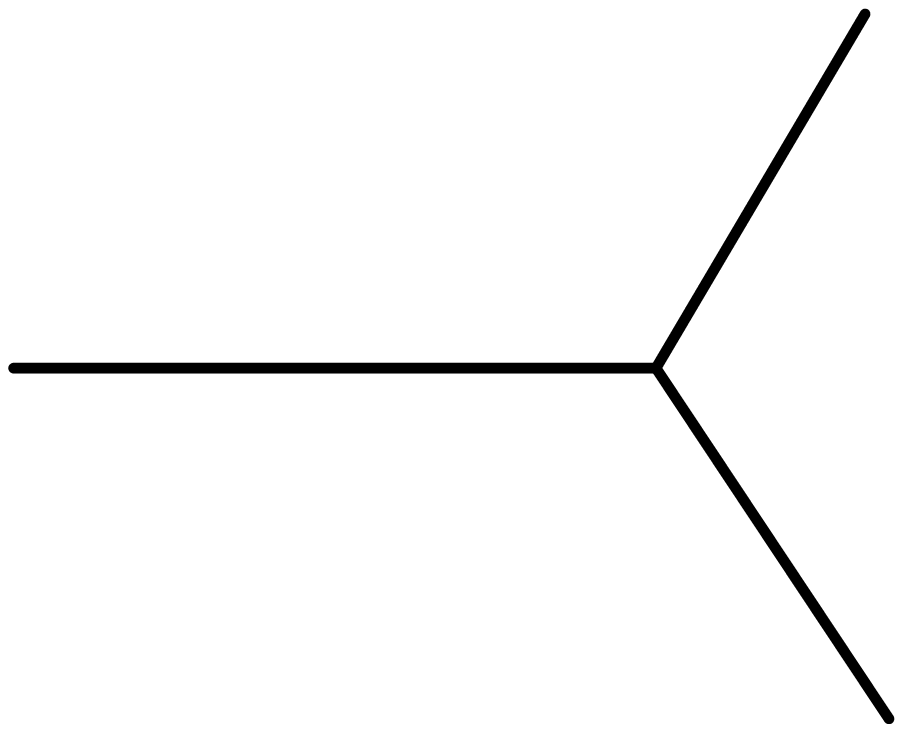}}}
\]
\caption{Left: 5d $\cN=2$ SYM on a junction of three cylinders. Right: The KK reduction leads to 4d $\cN=4$ SYM on three half-spaces, meeting on $\bR^3$.
\label{junction-5dSYM}}
\end{figure}

We can introduce a metric on $\cC$ such that the surface consists of three cylinders of circumference $\ell_{1,2,3}$ meeting smoothly at a junction, see figure \ref{junction-5dSYM}.
The behavior of the system at length scales $L$ far larger than $\ell_{1,2,3}$ is given by three segments of 4d $\cN=4$ SYM meeting at the same boundary $\bR^3$.
The action on each segment is \eqref{eq:4daction} with $T_i = \ell_i/g^2_{5d}$.
The boundary condition at the junction is half-BPS, because the original 5d SYM on $\cC$ is half-BPS.

Let us determine the boundary condition explicitly.
Classical configurations which contributes dominantly  to the path integral at the scale $L\gg \ell_{1,2,3}$ will have  $A,X,Y$ of order $L^{-1}$ and the action density should scale as $L^{-4}$.
Mark three $S^1_{(i)}$'s ($i=1,2,3$) very close to the junction as depicted in figure \ref{junction-5dSYM}, and call the region bounded by them as $S$.
When $L$ is very big, the non-liner term in the covariant derivative can be discarded compared  to the derivative,
and the dominant contribution to the action is
\begin{equation}
\sim \int_S d^5x \left[|dA|^2 + |dX_a|^2 +|\partial Y_a|^2\right] \;.
\end{equation}
The action density should be of order $L^{-4}$. Then in the large $L$ limit,
$Y_a$ need to be constant while $A$ and $X_a$ need to be flat.
We let $A^{(i)}$, $X_{1,2}^{(i)}$ and $Y_{1,2,3}^{(i)}$ be the values of $A$, $X_{1,2}$ and $Y_{1,2,3}$ on $S^1_{(i)}$.
The boundary condition for $Y$ is then given by
\begin{equation}
\vec Y^{(1)} = \vec Y^{(2)} = \vec Y^{(3)} \;.
\end{equation}
Flatness of $A$ translates to
\begin{equation}
\int_{S^1_{(1)}} A + \int_{S^1_{(2)}} A + \int_{S^1_{(3)}} A = 0
\end{equation}
giving $T_{1}A^{(1)} + T_{2}A^{(2)} + T_{3}A^{(3)}=0$ and similarly for $X_{1,2}$. Calling $A$ as $X_3$, we obtain
\begin{equation}
T_{1}\vec X^{(1)}+T_{2}\vec X^{(2)}+T_{3}\vec X^{(3)}=0 \;.
\end{equation}

The result agrees with what we deduced from the brane construction in \eqref{boo} and \eqref{zot}. However the derivation here has the merit that it is applicable also to the 6d $\cN=(2,0)$ theories of type D and E, for which we have not found a brane construction of the junction.

\subsection{$\cN=4$ SYM on a graph}

We found that 3d Sicilian theories can be engineered in a purely field theoretic way---without involving string theory anymore---by putting $\SU(N)$ $\cN=4$ SYM on a graph. The graph is made of segments, that can end on ``punctures'' or can be joined at trivalent vertices. On each segment we put a copy of $\SU(N)$ SYM. A puncture $\rho$ corresponds to the boundary condition $(\rho,1,\varnothing)$, while the trivalent vertex corresponds to the boundary condition $(1,\SU(N)^3, T_N)$.

To obtain the mirror theory we simply perform S-duality of $\cN=4$ SYM on each segment: $\SU(N)$ SYM is mapped to itself; the boundary conditions at the punctures are mapped to $\big( 1, \SU(N), T_\rho[\SU(N)] \big)$; the boundary condition at the vertices is mapped to $(1, \SU(N)_\text{diag}, \varnothing)$. To read off the 3d theory it is convenient to reduce the graph: every time we have SYM with breaking-to-the-diagonal vertices on both sides, the gauge group is broken, we can remove the segment and leave a $n$-valent vertex which breaks $\SU(N)^n$ to the diagonal $\SU(N)$.
If instead the two ends of the same segment are joined together, we are left with an adjoint hypermultiplet. This parallels the discussion of section~\ref{sec:mirror sicilian} and reproduces the star-shaped quivers.

\begin{figure}[t]
\[
\includegraphics[width=.95\textwidth]{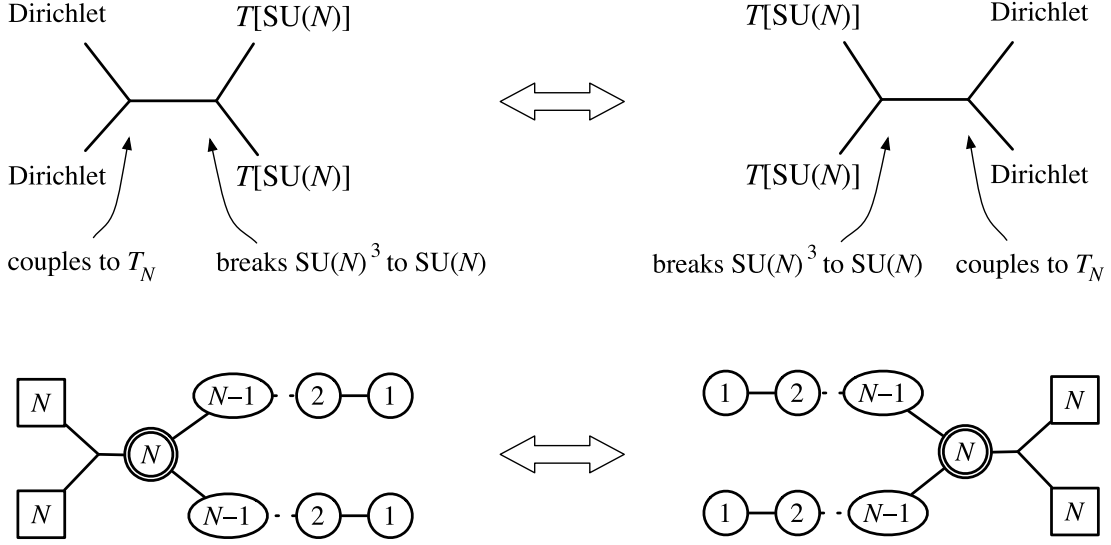}
\]
\caption{Mirror symmetry including both $T_N$ and star-quivers.
The 3d theory is given by a graph on which $\cN=4$ SYM lives. The mirror is obtained by performing the S-dual of the boundary conditions at open ends and at the junctions. The case depicted is self-mirror.
\label{mixed}}
\end{figure}

\begin{figure}[t]
\[
\includegraphics[width=.95\textwidth]{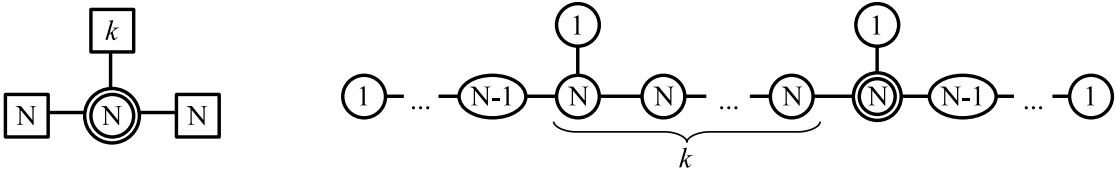}
\]
\vspace{-1.2cm}
\caption{Left: $k$ domain walls introducing one extra fundamental have been added, compared to figure \protect\ref{fig:mirror of SU3}. Right: Its mirror have $k$ domain walls, each introducing an extra bifundamental coupled to a $\U(1)$. 
\label{fig:extra fund}}
\end{figure}

The advantage of this perspective is that, being purely field theoretical, can be generalized beyond brane constructions.
For instance, we could couple star-shaped quivers to Sicilian theories: in this way we get a class of theories closed under mirror symmetry, see figure \ref{mixed}.
More generally, the full set of half-BPS boundary conditions in \cite{Gaiotto:2008sa,Gaiotto:2008ak} can be used. One example is a domain wall that introduces  a fundamental hypermultiplet; its mirror is a domain wall that introduces a bifundamental coupled to extra $\U(1)$, see figure \ref{fig:extra fund}.
Finally, one could consider $\cN=4$ SYM with gauge groups other than $\SU(N)$. We will consider $\SO(P)$ in the next section.

\section{$D_N$ Sicilian theories}
\label{sec:DN Sicilian}

Another class of Sicilian theories, that we will call of type $D_N$ and studied in \cite{Tachikawa:2009rb}, can be obtained by compactification of the 6d $\cN=(2,0)$ $D_N$ theory on a Riemann surface with half-BPS punctures. The 6d $D_N$ theory is the low energy theory on a stack of $2N$ $\frac12$M5-branes on top of the $\bR^5/\bZ_2$ orientifold in M-theory;
here and in the following, having $2N$ $\frac12$branes means to have $2N$ branes on the covering space.
This parallels the construction of Sicilian theories of type $A_{N-1}$ considered so far.
We are interested in extending the mirror  map to those theories.

Since the 6d $D_N$ theory compactified on $T^2$ gives 4d $\cN=4$ $\SO(2N)$ SYM at low energy, it should be possible to construct 3d $D_N$ Sicilian theories through $\cN=4$ SYM on a graph, with suitable half-BPS boundary conditions at the punctures and at the junctions. This is the approach we follow in this section.

\subsection{The punctures}

\begin{table}[t]
\begin{center}
\small{
\begin{tabular}{|c||c|c|c|c|}
\hline
O-plane & gauge theory & \tabs across $\frac12$D$(p+2)$ & across $\frac12$NS5 & S-dual ($p=3$) \\
\hline
\hline
O$p^-$  & O($2N$) & \tabs \Otm{$p$} & O$p^+$ & O3$^-$ \\
\hline
\tabs \Otm{$p$} &  O($2N+1$) & O$p^-$ & \Otp{$p$} & O3$^+$ \\
\hline
O$p^+$ & USp($2N$) & \tabs \Otp{$p$} & O$p^-$ & $\Otm{3}$ \\
\hline
\tabs \Otp{$p$} &  $\USp'(2N)$ & O$p^+$ & \Otm{$p$} & \Otp{3} \\
\hline
\end{tabular}
}
\caption{Properties of O$p$-planes, for $p \leq 5$. We indicate: type of O$p$-plane,  gauge theory living on them when $2N$ D$p$-branes are added to the covering space, type of O$p$-plane on the other side of a crossing $\frac12$D$(p+2)$-brane, or $\frac12$NS5-brane,
and S-dual plane (for $p=3$). In our conventions $\USp(2) \cong \SU(2)$, and $\USp'(2N)$ is $\USp(2N)$ with the $\theta$ angle shifted by $\pi$.
\label{tab:O-planes}}
\end{center}
\end{table}

Let us start by focusing on a single puncture, which can be understood via systems of D4/O4/D6-branes.  First consider the 6d $D_N$ theory on a cigar, with a single puncture at the tip, as we did for the $A_{N-1}$ theory in figure \ref{fig:puncture-5dSYM}.
Far from the tip we have the 6d $D_N$ theory on $S^1$, in other words
$2N$ $\frac12$D4-branes on top of an O4$^-$-plane.
The tip of the cigar with a puncture then becomes a half-BPS boundary condition for that theory,
which comes from terminating $\frac12$D4-branes on $\frac12$D6-branes. The configuration of branes is as in the case of $A_{N-1}$, see the table in figure \ref{fig:puncture-5dSYM}. In our conventions $\USp(2) \cong \SU(2)$.

Let us classify how $2N$ $\frac12$D4-branes on top of an O4$^-$-plane can end on $\frac12$D6-branes. Let us put as many D6-branes as possible away from the orientifold.
For each $\frac12$D6, assign a column of boxes whose height is given by the change in the D4-charge across the D6-brane.
We thus obtain Young diagrams with $2N$ boxes. Let $N_h$ be the number of columns of height $h$. When $N_h$ is even, we can place the $N_h$ $\frac12$D6-branes outside the O-plane, and no further restrictions apply. When $N_h$ is odd, one $\frac12$D6 has to be placed on top of the O-plane.  However, every time a $\frac12$D6 crosses the O4$^-$, the latter becomes an \Otm{4} on the other side, see \cite{Feng:2000eq} for more details.
Therefore the difference of the $\frac12$D4-charge is odd.
This implies that $N_h$ must be even for $h$ even.
We call these the \emph{positive} punctures, and the corresponding diagrams Young diagrams of $\O(2N)$.
The global symmetry algebra at these punctures is read off from the brane construction:
\be
\label{positive symmetry}
\fg_{\rho^+} = \bigoplus_{h \text{ odd}} \so(N_h) \oplus \bigoplus_{h \text{ even}} \usp(N_h) \;.
\ee

We also have \emph{negative} punctures, which produce a branch cut or twist line across which there is a $\bZ_2$ monodromy of the $D_N$ theory.%
\footnote{The 6d $D_N$ theory on a Riemann surface has operators of spin $2,4,\dots,2N-2$ plus one operator of spin $N$. They correspond to the Casimirs of $\so(2N)$, the last one being the Pfaffian. The $\bZ_2$ twist changes the sign of the operator of spin $N$, corresponding to the parity outer automorphism of $\so(2N)$.}
The monodromy will terminate on some other negative puncture on the Riemann surface. Compactifying the 6d $D_N$ theory on $S^1$ with such a twist, we obtain 5d $\cN=2$ $\USp(2N-2)$ SYM \cite{Vafa:1997mh}.
This time we have $(2N-2)$ $\frac12$D4-branes on top of an O4$^+$-plane.
The property of O4$^+$-planes crossing a $\frac12$D6-brane now implies that $N_h$ must be even when $h$ is odd, in contrast to the positive punctures.
We call these diagrams Young diagrams of $\USp(2N-2)$.
The global symmetry is now
\be
\label{negative symmetry}
\fg_{\rho^-} = \bigoplus_{h \text{ even}} \so(N_h) \oplus \bigoplus_{h \text{ odd}} \usp(N_h) \;.
\ee

The analysis here is equivalent to that given in  \cite{Tachikawa:2009rb}, except that we moved all the D6-branes to the far-right of the NS5-branes  and that we can thus read off the flavor symmetry.
So far we have considered the 6d $D_N$ theory on a cigar, which provides information about the 4d Sicilian theory; after compactification on $S^1$ we can perform a T-duality and repeat the whole construction in terms of D3/O3/D5-branes, which is useful to get the mirror.

The S-dual of the boundary conditions at the punctures are easily obtained from the brane construction, as written in \cite{Gaiotto:2008ak}.
We start with the brane setup of the puncture, given by $\frac12$D3-branes on top of an O3-plane and ending on $\frac12$D5-branes, and perform an S-duality transformation (table \ref{tab:O-planes}).
The resulting theory at the puncture is read off, recalling that $2k$ $\frac12$D3-branes on O3$^+$ or \Otp{3} and suspended between $\frac12$NS5-branes give an $\USp(2k)$ gauge theory, while $k$ $\frac12$D3-branes on O3$^-$ or \Otm{3} give an $\O(k)$ gauge theory.%
\footnote{At the level of the algebra, O3$^-$ and \Otm{3} project $\u(k)$ to its imaginary subalgebra which is $\so(k)$. At the level of the group, the projection selects the real subgroup of $\U(k)$, which is $\O(k)$.}

A positive puncture $\rho^+ = \{h_1, \dots, h_J\}$ before S-duality
describes D3-branes on an O3$^-$ puffing up to become D5-branes.
Accordingly, it should be given by an embedding $\rho^+: \SU(2) \to \SO(2N)$. Indeed, if we decompose the real $2N$-dimensional representation of $\SO(2N)$ in terms of irreducible representations of $\SU(2)$ as in \eqref{eq:decomp},
$N_h$ for even $h$  is even,  because $\underline{h}$ for even $h$ is pseudo-real.
The global symmetry \eqref{positive symmetry} is the commutant of this embedding $\rho^+$.
Performing S-duality and exchanging D5-branes with NS5-branes, we obtain the quiver
\be
\label{rho+ quiver}
\underline{\SO(2N)} - \USp(r_1) - \O(r_2) - \cdots - \USp(r_{J-1})
\ee
where the underlined group is a flavor Higgs symmetry as before. Here $J$ is always even, and the sizes are
\be
\label{rho+ ranks}
r_a = \Big[ \sum_{b=a+1}^J h_b \Big]_{+,-} \,,\qquad\qquad +: \O \;,\quad -: \USp
\ee
where $[n]_{+(-)}$ is the smallest (largest) even integer $\geq n$ ($\leq n$). The two options refer to the group being $\O$ or $\USp$.
When the last group is $\USp(0)$,  we  remove it. These quivers have been introduced in \cite{Gaiotto:2008ak} and called $T_{\rho^+}[\SO(2N)]$.

A negative puncture $\rho^-= \{h_1,\dots,h_J\}$ before S-duality
describes D3-branes on an O3$^+$ puffing up to become D5-branes.
Accordingly, it should be given by an  embedding $\rho^-: \SU(2) \to \USp(2N-2)$. Indeed, if we decompose the pseudo-real $(2N-2)$-dimensional representation of $\USp(2N-2)$ under $\SU(2)$,  $N_h$ for odd $h$ is even,  because  $\underline{h}$ is strictly  real when $h$ is odd.
The global symmetry \eqref{positive symmetry} is the commutant of this embedding $\rho^-$.
Performing S-duality,  we get the 3d quiver
\be
\label{rho- quiver}
\underline{\SO(2N-1)} - \USp(r_1) - \O(r_2) - \cdots - \O(r_{\tilde J}) \qquad\text{ with } \tilde J = [J]_+ \;.
\ee
The sizes are
\be
\label{rho- ranks}
r_a = \Big[ 1 + \sum_{b=a+1}^J h_b \Big]_{\widetilde+,-} \,,\qquad\qquad \widetilde+: \O \;,\quad -:\USp
\ee
where $[n]_{\widetilde+}$ is the smallest odd integer $\geq n$. The two options refer to the group being $\O$ or $\USp$. These quivers are called $T_{\rho^-}[\SO(2N-1)]$.

They give rise to the S-dual pairs of boundary conditions
\bea
& S:\; \big( \rho^+,\, 1,\, \varnothing \big)_{X,Y} \;\mapsto\; \big( 1,\, \SO(2N),\, T_{\rho^+}[\SO(2N)] \big)_{Y,X} \;, \\
& S:\; \big( \rho^-,\, 1,\, \varnothing \big)_{X,Y} \;\mapsto\; \big( 1,\, \O(2N-1),\, T_{\rho^-}[\SO(2N-1)] \big)_{Y,X} \;.
\eea

The Coulomb branch of $T_{\rho^+}[\SO(2N)]$ is  $\cS_{\rho^+} \cap \overline\cN\subset \so(2N)_\bC$, whereas that of $T_{\rho^-}[\SO(2N-1)]$ is $\cS_{\rho^-}\cap \overline\cN\subset \usp(2N-2)_\bC$.
As such, the symmetries on the Coulomb branch are given by the commutant of $\rho^+$ inside $\SO(2N)$ and of $\rho^-$ inside $\USp(2N-2)$, respectively. They agree with the symmetries found from the brane construction, \eqref{positive symmetry} and \eqref{negative symmetry}.
The theories $T_\rho[\SO(r)]$ have a Higgs branch which is the closure of a certain nilpotent orbit $\rho^\vee$ of $\O(r)$. We provide the algorithm to obtain $\rho^\vee$ in appendix \ref{app:rhovee}.

\subsection{Two types of junctions and their S-duals}

\begin{figure}
\[
\vcenter{\hbox{\includegraphics[scale=.4]{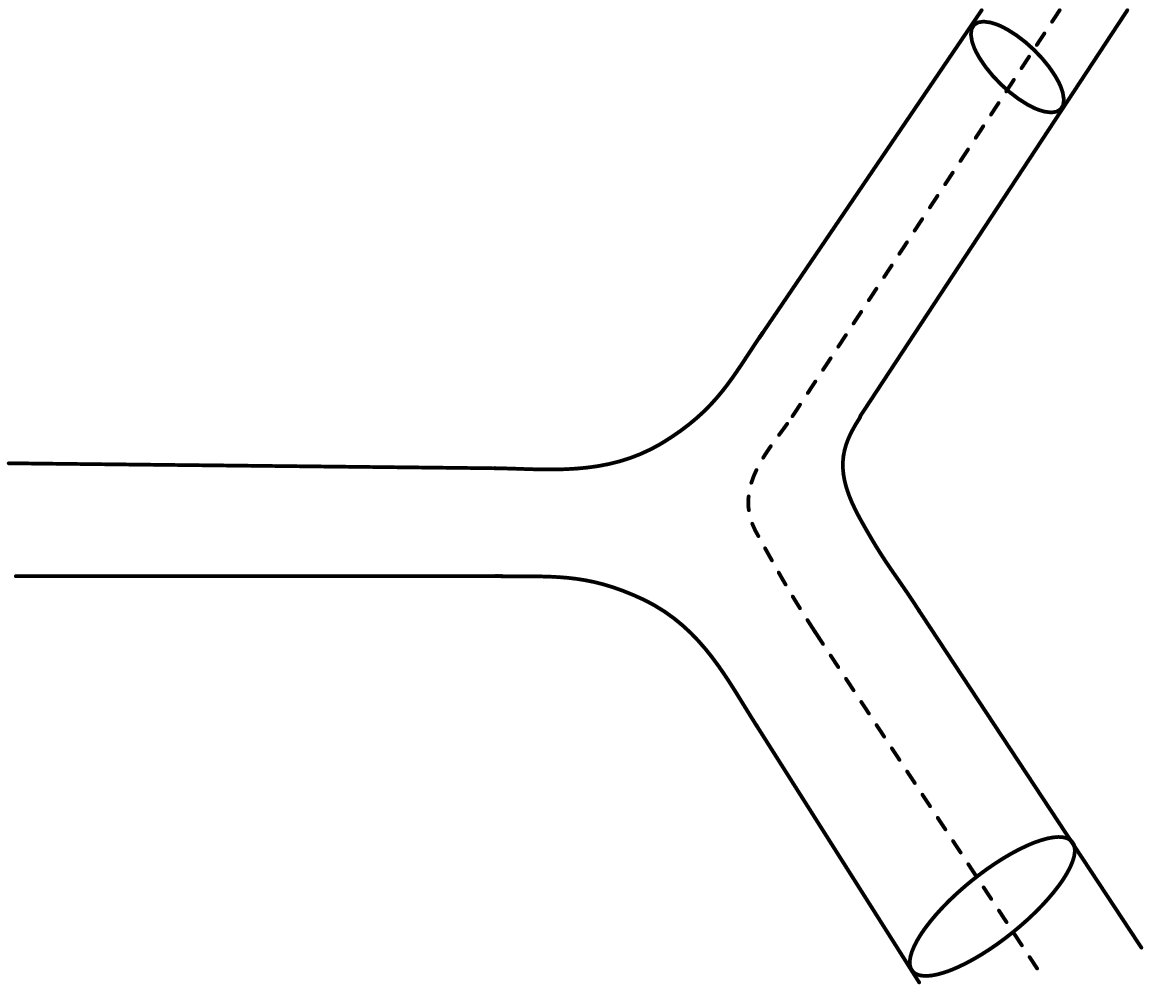}}} \qquad\Longrightarrow\qquad
\vcenter{\hbox{\includegraphics[scale=.4]{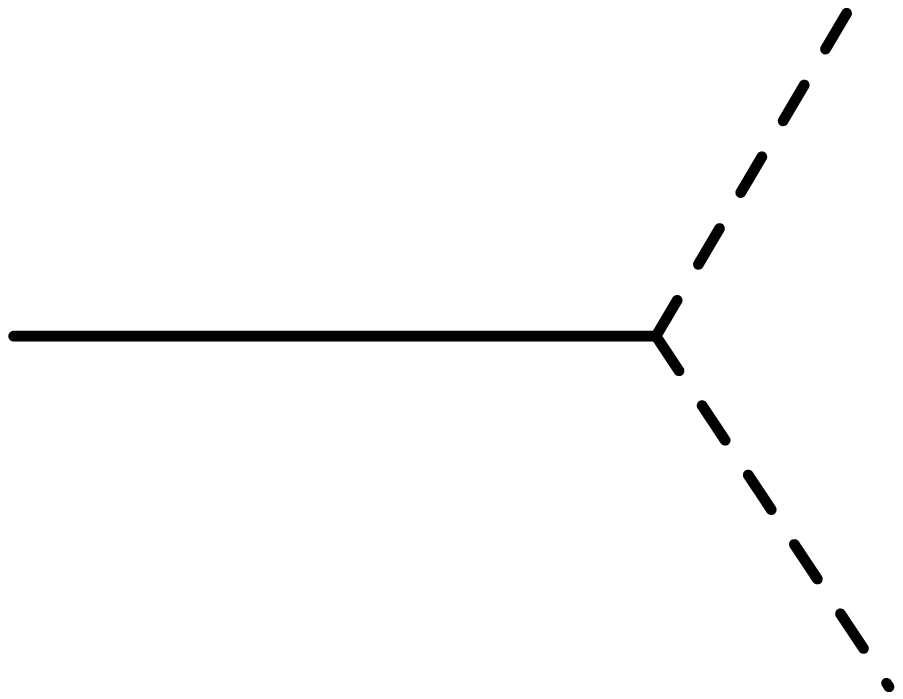}}}
\]
\caption{6d $D_N$ theory on a pair of pants $\cC$ with $\bZ_2$ monodromy along two tubes. KK reduction along $S^1_\cC \subset \cC$ gives a junction of two segments of 5d $\USp(2N-2)$ SYM and one of $\SO(2N)$; further compactification on $S^1_A$ gives its 4d version. Instead compactification on $S^1_A$ gives 5d $\SO(2N)$ SYM on a pair of pants with $\bZ_2$ monodromy by the parity transformation $\sigma \in \O(2N)$; further KK reduction on $S^1_\cC$ gives a junction of two segments of 4d $\O(2N-1)$ SYM and one of $\SO(2N)$.
We obtain $\USp(2N-2)$ in the first description and $\O(2N-1)$ in the second, because the procedure involves the exchange of $S^1_\cC$ and $S^1_A$, which acts as S-duality of 4d $\cN=4$ SYM.
\label{fig:dotted}}
\end{figure}

3d $D_N$ Sicilian theories can be constructed by putting $\cN=4$ SYM on a graph. We saw that there are two types of punctures: positive ones,
which are boundary conditions for $\SO(2N)$ SYM, and negative ones,
which are boundary conditions for $\USp(2N-2)$ SYM and create a twist line.
Accordingly, to  keep track of twist lines, on the segments of the graph we put either $\SO(2N)$ or $\USp(2N-2)$ SYM.
We need to consider two types of junctions: a junction among three copies of $\SO(2N)$ SYM, and a junction among one copy of $\SO(2N)$ and two copies of $\USp(2N-2)$. These junctions correspond to the maximal triskelions, see figure \ref{fig:dotted}.
We will call the two resulting theories $R_{2N}$ with $\SO(2N)^3$ Higgs symmetry
and $\widetilde R_{2N}$ with $\SO(2N) \times \USp(2N-2)^2$ Higgs symmetry.
When compactified on $S^1$, the boundary conditions at the junction are
\be
\label{untwisted junction BC}
\big( 1,\, \SO(2N)^3,\, R_{2N} \big)_{X,Y} \qquad\text{ and }\qquad \big( 1,\, \SO(2N)\times \USp(2N-2)^2,\, \widetilde R_{2N} \big)_{X,Y} \;.
\ee
With these boundary conditions, all 3d $D_N$ Sicilian theories can be reproduced via pants decomposition.

The S-dual of these boundary conditions can be easily obtained with the analysis in section~\ref{sec:junction}. We obtain
\bea
& S:\; \big( 1,\, \SO(2N)^3,\, R_{2N} \big)_{X,Y} &\;\mapsto\;& \big( 1,\, \SO(2N)_\text{diag},\, \varnothing \big)_{Y,X} \,,\\
& S:\; \big( 1,\, \SO(2N) \times \USp(2N-2)^2,\, \widetilde R_{2N} \big)_{X,Y} &\;\mapsto\; &\big( 1,\, \O(2N-1)_\text{diag},\, \varnothing \big)_{Y,X} \;.
\eea
Here $\SO(2N)_\text{diag}$ is the diagonal subgroup of $\SO(2N)^3$, while $\O(2N-1)_\text{diag} \subset \SO(2N) \times \O(2N-1)^2$ corresponds to choosing an $\O(2N-1)$ subgroup of $\SO(2N)$, and then taking the diagonal subgroup of $\O(2N-1)^3$.

This can be proved also by considering a simple case in which we already know the mirror symmetry map.
For instance, consider the 3d Sicilian theory given by one simple positive puncture on the torus: this is 3d $\cN=8$ $\SO(2N)$ SYM. The graph construction has two $\SO(2N)$ segments, $(1,\SO(2N)^3,R_{2N})$ at the junction and $(\{2N-1,1\},1,\varnothing)$ at the puncture. The mirror theory is $\cN=8$ $\SO(2N)$ SYM itself. The S-dual graph has $\SO(2N)$ on the segments and $(1,\SO(2N),\varnothing)$ at the puncture, because $T_{\{2N-1,1\}}[\SO(2N)]$ is an empty theory. To reproduce the mirror, we need $(1,\SO(2N)_\text{diag},\varnothing)$ at the junction.
Similarly, consider the 3d Sicilian theory given by one simple positive puncture on the torus with a twist line around it: this is 3d $\cN=8$ $\USp(2N-2)$ SYM. The graph construction has one $\SO(2N)$ and one closed $\USp(2N-2)$ segment, $(1,\SO(2N)\times \USp(2N-2)^2, \widetilde R_{2N})$ at the junction and $(\{2N-1,1\},1,\varnothing)$ at the puncture. The mirror theory is $\cN=8$ $\O(2N-1)$ SYM. The S-dual graph has $\SO(2N)$ and $\O(2N-1)$ on the segments, and $(1,\SO(2N),\varnothing)$ at the puncture. To reproduce the mirror, we need $(1,\O(2N-1)_\text{diag},\varnothing)$ at the junction.

\subsection{Mirror of Sicilian theories}

Now it is easy to construct the mirrors of 3d Sicilian theories of type $D_N$
obtained from an arbitrary punctured Riemann surface $\cC$.
First consider $\cC$ of genus zero with only positive punctures.
When we gauge two $T[\SO(2N)]$ together via their $\SO(2N)$ Coulomb symmetries,
the Higgs branch of the combined theory is the cotangent bundle $T^*\SO(2N)_\bC$,
which has the action of $\SO(2N)\times \SO(2N)$ from the left and the right.
This is broken to its diagonal subgroup on the zero-section.
When it is gauged on both sides by different vector multiplets, the Higgs mechanism gets rid of one $\SO(2N)$ vector and the adjoint hypermultiplet. We are left with a star-shaped quiver with $\SO(2N)$ gauge group at the center. Then consider $\cC$ with genus $g\geq 1$ and with only positive punctures.
There will be $g$ copies of $T^*\SO(2N)_\bC$ gauged on both sides by the same $G=\SO(2N)$, \ie{} $G$ acts on the cotangent bundle by the adjoint action.
Around the origin of $T^*\SO(2N)_\bC$ all hypermultiplets are massless. We are left with a star-shaped quiver, with $g$ extra $\SO(2N)$ adjoint hypermultiplets and $\SO(2N)$ gauge group at the center.
The analysis so far was completely parallel to that of type $A_{N-1}$ Sicilians.

\begin{figure}
\[
\begin{array}{cc@{\hspace{3ex}}cc}
\hbox{a)}& \vcenter{\hbox{\includegraphics[scale=.6]{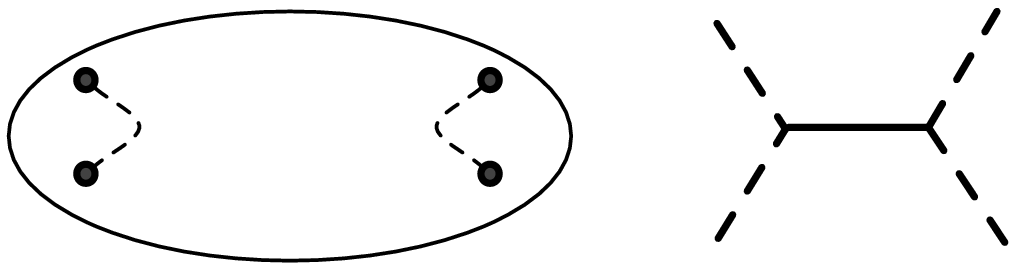}}} &
\hbox{b)}& \vcenter{\hbox{\includegraphics[scale=.6]{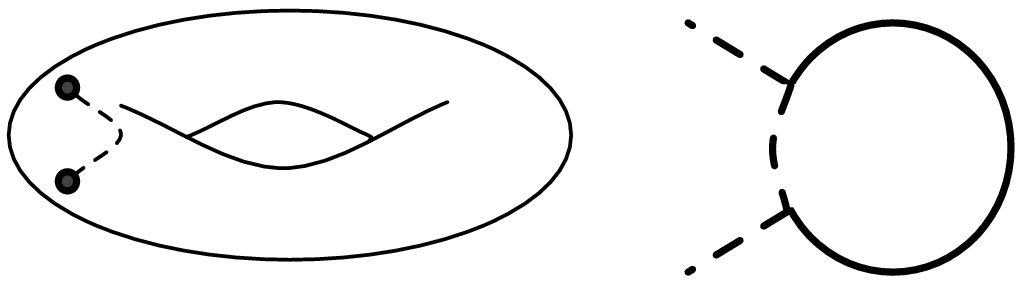}}} \\
\\
\hbox{c)}& \vcenter{\hbox{\includegraphics[scale=.6]{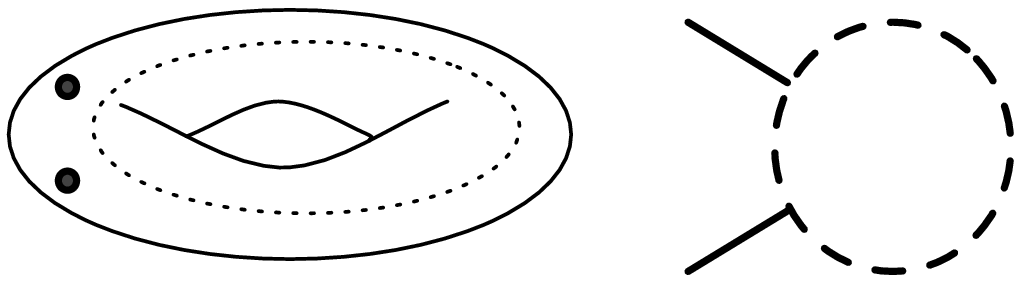}}} &
\hbox{d)}& \vcenter{\hbox{\includegraphics[scale=.6]{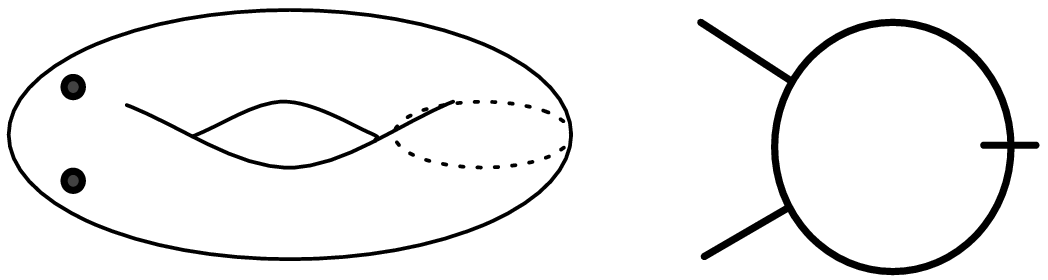}}} \\
\end{array}
\]
\caption{a) The segment at the center gives $T^*\SO(2N)_\bC$, to which $\O(2N-1)^2$ gauge groups couple. We are left with one $\O(2N-1)$ gauge group and one extra fundamental hypermultiplet.
b) $T^*\SO(2N)_\bC$  is coupled to $\O(2N-1)$ which acts by the adjoint action. We are left with an adjoint and a fundamental of $\SO(2N-1)$.
c) With a monodromy when one crosses a big $S^1$, the resulting 4d SYM on a graph has segments with $\O(2N-1)$ gauge group.
d) A monodromy when one crosses a small $S^1$ results
in 4d SYM on a graph with a loop around which we have a monodromy. This is indicated as a mark in the graph shown on the right.  c) and d) give rise to the same 3d theory in the low energy limit, as explained in the main text.
\label{fig:Dglue}}
\end{figure}

Next, consider $\cC$ of genus zero with $n_+$ positive and $2n_-$ negative punctures.
When we gauge together two copies of $T[\SO(2N-1)]$ on the Coulomb branch,
we get $T^*\SO(2N-1)_\bC$ which spontaneously breaks the symmetry.
However there will be $n_- -1$ copies of $T^*\SO(2N)_\bC$ which are gauged by two $\O(2N-1)$ on both sides: the gauge group is broken to the diagonal $\O(2N-1)$ and a hypermultiplet in the fundamental of $\O(2N-1)$ remains massless.
We are left with a star-shaped quiver, with $n_- - 1$ extra fundamentals of the $\O(2N-1)$ gauge group at the center. See figure \ref{fig:Dglue}a for the case $n_-=2$.

When $g>0$, there are many choices for the configuration of monodromies: they are classified by $H_1(\cC\setminus\{\text{punctures}\},\bZ_2)$.
When one has two negative punctures but without twist lines on a handle (see figure \ref{fig:Dglue}b), $T^*\SO(2N)_\bC$ is gauged by the same $\O(2N-1)$ via the adjoint action and we get one adjoint and one fundamental of $\SO(2N-1)$.
Another possibility is to have a closed twist line along a handle of the graph (figure~\ref{fig:Dglue}c): $T^*\SO(2N-1)_\bC$ is gauged on both sides by the same $G=\O(2N-1)$ via the adjoint action, giving rise to an adjoint of $\SO(2N-1)$.
If we take the S-duality of the 4d Sicilian theory first and then compactify it down to 3d, we get the 4d SYM on a graph shown in figure \ref{fig:Dglue}d.
This amounts to gauging $T^*\SO(2N)_\bC$ with one $\SO(2N)$ with the embedding
\begin{equation}
\SO(2N) \ni g \mapsto (g, \sigma g \sigma) \in \SO(2N)\times\SO(2N)
\end{equation}
where $\sigma \in \O(2N)$ is the parity transformation.
The theory spontaneously breaks the gauge group to $\O(2N-1)$, which is the subgroup of $\SO(2N)$ invariant under parity,
and eats up $2N-1$ hypermultiplets.
We are left with $\O(2N-1)$ with just one adjoint.

\begin{figure}[t]
\begin{center}
\includegraphics[width=.95\textwidth]{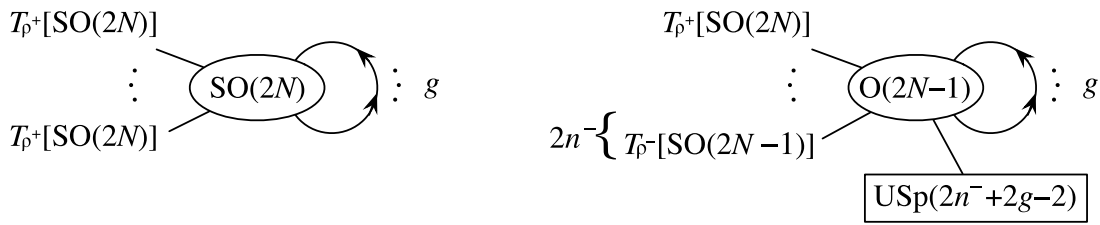}
\caption{Mirrors of $D_N$ Sicilian theories of genus $g$. Left: Mirror of a $D_N$ Sicilian with only positive punctures and no twist lines. The $g$ adjoints carry $\USp(2g)$ Higgs symmetry. Right: Mirror in the presence of some twist line, where $2n^- \geq 0$ is the number of negative punctures. The $n^- + g-1$ extra fundamentals carry $\USp(2n^- + 2g-2)$ Higgs symmetry.
\label{fig:mirror DN}}
\end{center}
\end{figure}

Summarizing, consider a 3d Sicilian theory defined by a genus $g$ Riemann surface $\cC$, some number of punctures (of which $2n_-$ are negative) and possibly extra closed twist lines in $H^1(\cC,\bZ_2)$. If there are no twist lines at all (so $n_-=0$), the mirror is a star-shaped quiver where an $\SO(2N)$ group gauges together all positive punctures and $g$ extra adjoint hypermultiplets. If there are twist lines, the mirror is a star-shaped quiver where an $\O(2N-1)$ group gauges together all punctures, $g$ extra adjoints and $(n_- + g - 1)$ extra fundamentals. This is summarized in figure \ref{fig:mirror DN}.
It is reassuring to find that the resulting mirror theory does not depend on the pants decomposition.

\section{Discussion}
\label{sec:discussion}

In this paper we have presented 3d theories, mirror to the $\cN=4$ 3d Sicilian theories of type $A_{N-1}$ and $D_N$. Although the latter do not have a simple Lagrangian description, the former turned out to be standard quiver gauge theories. On the way we have introduced a purely field theoretical construction to engineer 3d $\cN=4$ field theories: $\cN=4$ SYM on a graph. One considers a graph of segments and trivalent vertices, with half-BPS boundary conditions at the junctions and at the open ends.
Such a framework allows to formulate mirror symmetry as a ``modular'' operation: for 3d $\cN=4$ theories that can be decomposed in this way, we can perform mirror symmetry on the constituent blocks and finally put them together.

There are a number of directions in which this can be generalized, and many properties of 3d Sicilian theories to be explored.
For instance,  6d $\cN=(2,0)$ theories are characterized by their ADE type, but  in the language of $\cN=4$ SYM on a graph nothing prevents us from considering generic gauge groups. It would be interesting to work out the mirror symmetry map in those cases.
Along the same lines, the set of boundary conditions can be enlarged. Sticking with half-BPS boundary conditions with a brane realization, for which it is easy to take the S-dual, we could put together D5- and NS5-branes, and orbifolds/O5-planes, along the lines of \cite{Gaiotto:2008ak}. This would correspond to considering new tails in the quiver.

One of the most attracting directions is the addition of Chern-Simons terms and their behavior under mirror symmetry. So far Abelian CS terms in the context of mirror symmetry have been considered in \cite{Kapustin:1999ha}. It would be interesting to understand non-Abelian CS terms. They can be introduced from both sides of the mirror symmetry, and have different brane realizations.

Finally, it would be nice to generalize the construction of this paper to lower supersymmetry, for instance to 3d $\cN=2$ theories. In \cite{Benini:2009mz} 4d $\cN=1$ Sicilian theories have been constructed, of which one can consider the compactification. The particular mass deformation considered in that work has a brane realization in terms of D3-branes suspended between rotated NS5-branes.
As such it is possible to take the S-dual, giving the mirror in three dimensions.

\acknowledgments

The authors thank Davide Gaiotto and Brian Wecht for participation at an early stage of this work. FB thanks the Aspen Center for Physics where this work was completed.
The work of FB is supported in part by the US NSF under Grants No. PHY-0844827 and PHY-0756966. YT is supported in part by the NSF grant PHY-0503584, and by the Marvin L. Goldberger membership at the Institute for Advanced Study.

\appendix

\section{Hitchin systems and mirror quivers}
\label{app:Hitchin}

We perform in this appendix a stronger check of the proposed mirror symmetry map between 3d Sicilian theories and star-shaped quivers. We show that the Coulomb branch of the Sicilian theory is equal, as a hyperk\"ahler manifold, to the Higgs branch of the star-shaped quiver. It would be interesting to show the opposite.

A 4d Sicilian theory is defined by the compactification of $N$ M5-branes on a Riemann surface $\cC$ with punctures $p_i$ marked by Young diagrams $\rho_i$. Then its $S^1$ compactification gives
$N$ D4-branes on $\cC$ with the same punctures $p_i$ labeled by $\rho_i$. The Coulomb branch of this theory is the moduli space of Hitchin's equation given by such data. The interplay of the 4d $\cN=2$ theory compactified on $S^1$ and the Hitchin system was pioneered by Kapustin \cite{Kapustin:1998xn}, and was studied in great detail in \cite{Gaiotto:2009hg}. Also see \cite{Nanopoulos:2009uw}.

We show that in the low energy limit it coincides with the Higgs branch of its mirror quiver, which is easily computed via the F-term equations.
The relation between the moduli space of the Hitchin system and star-shaped quivers was studied by Boalch \cite{Boalch}.

\subsection{Hitchin systems}

Recall the nilpotent orbit $\overline\cN_{\rho}$,
which is the Higgs branch of $T_{\rho^\trans }[\SU(N)]$.
This is a hyperk\"ahler cone with triholomorphic $\SU(N)$ symmetry.
As such, it has a triholomorphic moment map $(\mu_\bR,\mu_\bC)$
\begin{equation}
\mu_\bR: \overline\cN_\rho \,\to\, \mathfrak{su}(N) \;, \qquad\qquad
\mu_\bC: \overline\cN_\rho \,\to\, \mathfrak{sl}(N) \;,
\end{equation}
where $\mathfrak{su}(N)$ and $\mathfrak{sl}(N)$ are Lie algebras.
One notable feature is that $\mu_\bC$ is just the embedding of $\overline\cN_\rho$ as complex matrices into $\mathfrak{sl}(N)$.

The Hitchin equation is a coupled differential equation for an $\SU(N)$ connection $A$ and a complex adjoint-valued $(1,0)$-form $\varphi$ on $\cC$; in terms of the variables in section \ref{sec:junction}, $\varphi=X_1+iX_2$.
We also have  degrees of freedom localized at the punctures $p_i$, parameterizing the nilpotent orbits $\overline\cN_{\rho_i^\trans }$.
The equations are given by
\begin{align}
F+ [\varphi,\bar\varphi] &=2 \sum_i\mu_{\bR}^{(i)} \, \delta(z-z_i) \;, \label{HitchinR} \\
\bar\partial_A \varphi &=\phantom{2} \sum_i\mu_{\bC}^{(i)} \, \delta(z-z_i) \;. \label{HitchinC}
\end{align}
Here $F$ is the curvature of $A$, $\bar\partial_A$ is the covariant holomorphic exterior derivative,
and $\mu_{\bR,\bC}^{(i)}$ comprise the triholomorphic moment map of $\overline\cN_{\rho_i^\trans }$.
To be precise, the term $[\varphi,\varphi^\dagger]$ should carry a factor of the radius of $S^1$ on which the 4d theory is compactified. However our 4d theory is superconformal, therefore the radius of $S^1$ is the only scale which we fix to one.

The space of solutions, with identification by $\SU(N)$ gauge transformations, is the moduli space $\cM$ of the Hitchin equation.
This set of equations is an infinite-dimensional version of hyperk\"ahler reduction.

\subsection{IR limit}

As stressed above, the 4d theory is conformal but its $S^1$ compactification is not because the radius of $S^1$ introduces a scale. The IR limit corresponds to measuring the system in a far larger distance scale compared to the radius. In terms of the moduli space, one needs to take a point on the moduli space and zoom in.

The point we are interested in is the origin of the moduli space, where the Coulomb and Higgs branch meet, which is at $A = \varphi = \mu^{(i)}_{\bR,\bC} = 0$.
The metric at the origin of the space of $A$, $\varphi$, $\mu^{(i)}_{\bR,\bC}$ is invariant under the scaling \begin{equation}
A\to \epsilon A, \quad
\varphi\to \epsilon \varphi, \quad
\mu^{(i)}_{\bR,\bC} \to \epsilon^2 \mu^{(i)}_{\bR,\bC} \;.
\end{equation} This scaling is inherited under the infinite-dimensional hyperk\"ahler reduction which gives the Hitchin equations.
Therefore we perform the expansion
\bea
\label{expansion}
A_\text{full} &= \epsilon A_1 + \epsilon^2 A_2 + \cdots \,,\\
\varphi_\text{full} &= \epsilon \varphi_1 + \epsilon^2 \varphi_2 + \cdots \,,\\
\mu_{\bR,\bC}^{(i)}{}_\text{full} &= \phantom{\epsilon \varphi_1 +}\ \epsilon^2 \mu_{\bR,\bC}^{(i)} \;.
\eea
and substitute them into \eqref{HitchinR}-\eqref{HitchinC}.

At order $\epsilon$, Eq.~\eqref{HitchinR} implies that $A_1$ is closed, and we
fix the gauge by demanding $A_1$ to be harmonic:
therefore $A_1$ has $2g(N^2-1)$ real degrees of freedom.
Eq.~\eqref{HitchinC} implies $\bar\partial \varphi_1=0$:
therefore $\varphi_1$ has $g(N^2-1)$ complex degrees of freedom.
It is convenient to package these degrees of freedom by expanding them as
\begin{equation}
A_1 = P_a \omega^a + P^\dagger_a \bar\omega^a \;, \qquad\qquad
\varphi_1 = Q_a \omega^a \;,
\end{equation}
where $\omega^a$ ($a=1,\ldots,g$) is the basis of holomorphic $(1,0)$-forms satisfying
\begin{equation}
\int \omega^a \wedge \bar\omega^b = \delta^{ab} \;.
\end{equation}

At order $\epsilon^2$ the Hitchin equation reads
\bea
dA_2 + A_1\wedge A_1 +[\varphi_1,\bar\varphi_1] &= 2\sum_i \mu_{\bR}^{(i)} \delta(z-z_i) \;, \\
\bar\partial\varphi_2 + A_1\wedge \varphi_1 &= \phantom{2} \sum_i \mu_{\bC}^{(i)}\delta(z-z_i) \;.
\eea
This can be solved if and only if
\begin{equation}
\int \bigl(A_1\wedge A_1 +[\varphi_1,\bar\varphi_1] \bigr) = 2\sum_i \mu_{\bR}^{(i)} \;, \qquad\qquad
\int A_1\wedge \varphi_1 = \sum_i \mu_{\bC}^{(i)} \;.
\end{equation}
In terms of $P_i$ and $Q_i$, we find
\bea
\label{hk}
0 &= 2\sum_i \mu_{\bR}^{(i)} - \sum_a \bigl([P_a,P_a^\dagger] - [Q_a,Q_a^\dagger]\bigr) \;, \\
0 &= \sum_i \mu_{\bC}^{(i)} - \sum_a [P_a,Q_a] \;.
\eea

Assuming that the higher-order terms  in the expansion \eqref{expansion} can be recursively defined, the near-origin moduli space $\cM_0$
is obtained by identifying the solutions of \eqref{hk} by $\U(N)$ conjugation.
This space is exactly the hyperk\"ahler quotient
\begin{equation}
\cM_0 = \Bigl[\underbrace{V\times V\times \cdots \times V}_\text{$g$ times}\times  \overline\cN_{\rho^\trans _1}\times\overline\cN_{\rho^\trans _2}\times \cdots \Bigr] \hkq \SU(N) \;,
\end{equation}
where $V=\su(N)_\bC\oplus\su(N)_\bC$ is the space of two $N\times N$ traceless matrices $P$ and $Q$.

Recalling that $\cN_{\rho^\trans }$ is precisely the Higgs branch of the quiver theory $T_{\rho}[\SU(N)]$, we conclude that $\cM_0$ is the Higgs branch of the theory consisting of the collection of $T_{\rho_i}[\SU(N)]$, for all $i$, and $g$ extra adjoint hypermultiplets, coupled to a $\U(N)$ vector multiplet.
These are exactly the star-shaped quiver theories which we argued to be the mirror of our theory.

\section{More on S-dual of punctures of type $D$}
\label{app:rhovee}

Here we explicitly identify the Higgs branch of $T_\rho[\SO(P)]$
as a nilpotent orbit $\rho^\vee$ of $\O(P)_\bC$, using the analysis in \cite{Kobak:2001}.
This was not explicitly written down in \cite{Gaiotto:2008ak}.

In 3d $\cN=2$ notation, the matter content  of the quiver $T_\rho[\SO(P)]$, defined by $\rho = \{h_1,\dots,h_J\}$ as in (\ref{rho+ quiver}) \dots (\ref{rho- ranks}), are given by adjoints $\Phi_i$ of size $r_i\times r_i$ and bifundamental chiral superfields $C_i$ which are  $r_{i-1} \times r_i$ complex matrices.   Let $Q$ be the number of the gauge groups.
The representations of $\O(r_i)$ are real, while indices of $\USp(r_i)$ are conjugated with $\eta$, the antisymmetric invariant tensor of USp.
The superpotential is
\be
W = \sum_{\substack{i=1 \\ i \text{ odd}}}^Q \Phi_i \eta \big( C_i^\trans C_i - C_{i+1} C^\trans_{i+1} \big) \eta + \sum_{\substack{i=2 \\ i \text{ even}}}^Q \Phi_i \big( C_i^\trans \eta C_i - C_{i+1} \eta C_{i+1}^\trans \big)
\ee
where we set $C_i \equiv 0$ for $i>Q$.
The F-term equations are (on the Higgs branch $\Phi_i = 0$):
\bea
C_i^\trans C_i &= C_{i+1} C_{i+1}^\trans \qquad & \text{for } & i \ \text{odd}, \\
C_i^\trans \eta C_i &= C_{i+1} \eta C_{i+1}^\trans & \text{for } &i \ \text{even} \;.
\eea

We construct the gauge invariant quantity $M = C_1 \eta C_1^\trans$, which is an element of $\so(P)_\bC$: in fact it parameterizes the Higgs branch and it is the moment map of $\SO(P)$ on it.
From the F-term equations and noticing that $C_i$ are rectangular matrices, we get:
\be
\label{rank equation}
\rk M^i \leq r_i \qquad \forall\; i = 1,\dots, Q \;, \qquad\qquad M^{Q+1} = 0 \;.
\ee
These equations define a set of nilpotent matrices. Notice that not always a nilpotent matrix can saturate the inequalities in (\ref{rank equation}): for any matrix $M$, the ranks $\tilde r_i = \rk M^i$ ($i = 0,\dots$) are such that $\tilde r_i - \tilde r_{i-1}$ is a non-increasing function of $i$.
In fact, $r_i$ are such that the solutions to (\ref{rank equation}) are matrices with ranks $\tilde r_i$ which are the largest integers $\tilde r_i \leq r_i$ with the latter property. Antisymmetry of $M$ does not impose any further constraint.

For $P = 2N$, (\ref{rank equation}) defines the closure of a nilpotent orbit $\cN_{\rho^{+\vee}}$ of $\O(2N)_\bC$, where $\rho^{+\vee}$ is a Young diagram of $\O(2N)$. It can be constructed with the following algorithm. Start with $\rho^+$, a Young diagram of $\O(2N)$. Take its transpose $\rho^{+\trans} = \{l_1 ,\dots, l_K\}$, where $l_1 \geq \dots \geq l_K$ are the lengths of the rows of $\rho^+$ and $K = h_1$.
Note that in general $\rho^{+\trans}$ is not a diagram of $\O(2N)$.
For every $i$, starting from 1 to $K$, we perform the following operation: If $l_i$ is even and $N_{l_i}$ is odd so that the diagram violates the rule of $\O(2N)$,
we let $\tilde\jmath$ be the largest $j$ such that $l_j=l_i$ and reduce $l_{\tilde\jmath} \to l_{\tilde\jmath} - 1$.
Then we let $\tilde k$ be the smallest $k$ such that $l_k \leq l_i-2$, and increase $l_{\tilde k} \to l_{\tilde k} + 1$. We then proceed in the algorithm with $i+1$.
The map $\rho^+ \to \rho^{+\vee}$ from the set of Young diagrams of $\O(2N)$ to itself is, in general, neither surjective nor injective. In particular it is not an involution.

For $P = 2N-1$, (\ref{rank equation}) defines the closure of a nilpotent orbit $\cN_{\rho^{-\vee}}$ of $\O(2N-1)_\bC$, where $\rho^{-\vee}$ is a Young diagram of $\O(2N-1)$.
This time the algorithm is the following. Start with $\rho^-$, a Young diagram of $\USp(2N-2)$. Take its transpose $\rho^{-\trans} = \{l_1,\dots,l_K\}$ and then sum 1 to the first length: $\{l_1 + 1, l_2,\dots,l_K\}$. In general this is not a diagram of $\O(2N-1)$. Finally perform the same algorithm as before. Again, the map $\rho^- \to \rho^{-\vee}$ from the set of Young diagrams of $\USp(2N-2)$ to those of $\O(2N-1)$ is neither surjective nor injective, and not an involution.

\linespread{1}
\bibliography{bib}{}
\bibliographystyle{utphys}

\end{document}